\documentstyle[aps,prd,psfig,floats,amssymb,times,colordvi]{revtex}
\begin{document}
\input{epsf.sty}

\twocolumn[\hsize\textwidth\columnwidth\hsize\csname @twocolumnfalse\endcsname
 
\title{Three-dimensional general relativistic
hydrodynamics II: long-term dynamics of
\newline
single relativistic stars}

\author{
Jos{\'e} A. Font${}^{(1,2)}$,
Tom Goodale${}^{(3)}$,
Sai Iyer${}^{(4)}$,
Mark Miller${}^{(4)}$,
Luciano Rezzolla${}^{(5,6)}$,
Edward Seidel${}^{(3,7)}$,
\newline
Nikolaos Stergioulas${}^{(8)}$, 
Wai-Mo Suen${}^{(4,9)}$,
Malcolm Tobias${}^{(4)}$ \medskip
}

\address{
$^{(1)}$Max-Planck-Institut f{\"u}r Astrophysik,
Karl-Schwarzschild-Str. 1, D-85740 Garching, Germany            \\
$^{(2)}$Departamento de Astronom\'{\i}a y Astrof\'{\i}sica,
Universidad de Valencia, 46100 Burjassot (Valencia), Spain      \\
$^{(3)}$Max-Planck-Institut f{\"u}r Gravitationsphysik,
Am M\"{u}hlenberg 1, D-14476 Golm, Germany                      \\
$^{(4)}$McDonnell Center for the Space Sciences, Department of Physics,
Washington University, St. Louis, Missouri, 63130               \\
$^{(5)}$SISSA, International School for Advanced Studies,
Via Beirut 2-4, 34014 Trieste, Italy                            \\
$^{(6)}$INFN, Department of Physics, University of
	Trieste, Via A. Valerio 2, 34127 Trieste, Italy		\\
$^{(7)}$National Center for Supercomputing Applications,
Beckman Institute, 405 N. Mathews Ave., Urbana, IL 61801        \\
$^{(8)}$Department of Physics, Aristotle University of Thessaloniki,
	Thessaloniki 54006, Greece				\\
$^{(9)}$Physics Department, Chinese
	University of Hong Kong, Shatin, Hong Kong
}
\date{\today}

\maketitle

\begin{abstract}                                                        %
 
This is the second in a series of papers on the
construction and validation of a three-dimensional code
for the solution of the coupled system of the Einstein
equations and of the general relativistic hydrodynamic
equations, and on the application of this code to
problems in general relativistic astrophysics. In
particular, we report on the accuracy of our code in the
long-term dynamical evolution of relativistic stars and
on some new physics results obtained in the process of
code testing. The following aspects of our code have been
validated: the generation of initial data representing
perturbed general relativistic polytropic models (both
rotating and non-rotating), the long-term evolution of
relativistic stellar models, and the coupling of our
evolution code to analysis modules providing, for
instance, the detection of apparent horizons or the
extraction of gravitational waveforms. The tests involve
single non-rotating stars in stable equilibrium,
non-rotating stars undergoing radial and quadrupolar
oscillations, non-rotating stars on the unstable branch
of the equilibrium configurations migrating to the stable
branch, non-rotating stars undergoing gravitational
collapse to a black hole, and rapidly rotating stars in
stable equilibrium and undergoing quasi-radial
oscillations. We have carried out evolutions in full
general relativity and compared the results to those
obtained either with perturbation techniques, or with
lower dimensional numerical codes, or in the Cowling
approximation (in which all the perturbations of the
spacetime are neglected). In all cases an excellent
agreement has been found. The numerical evolutions have
been carried out using different types of polytropic
equations of state using either the rest-mass density
only, or the rest-mass density and the internal energy as
independent variables. New variants of the spacetime
evolution and new high resolution shock capturing (HRSC)
treatments based on Riemann solvers and slope limiters
have been implemented and the results compared with those
obtained from previous methods. In particular, we have
found the ``monotonized central differencing'' (MC)
limiter to be particularly effective in evolving the
relativistic stellar models considered. Finally, we have
obtained the first eigenfrequencies of rotating stars in
full general relativity and rapid rotation. A long
standing problem, such frequencies have not been obtained
by other methods. Overall, and to the best of our
knowledge, the results presented in this paper represent
the most accurate long-term three-dimensional evolutions
of relativistic stars available to date.

 \end{abstract}


]

\section{Introduction}                                                  %
\label{introduction}                                                    %

        Computational general relativistic astrophysics
is an increasingly important field of research. Its
development is being driven by a number of
factors. Firstly, the large amount of observational data
by high-energy X-ray and $\gamma$-ray satellites such as
Chandra, XMM and others~\cite{nasa_missions}. Secondly,
the new generation of gravitational wave detectors coming
online in the next few years~\cite{Thorne94a}, and
thirdly, the rapid increase in computing power through
massively parallel supercomputers and the associated
advance in software technologies, which make large-scale,
multi-dimensional numerical simulations
possible. Three-dimensional (3D) simulations of general
relativistic astrophysical events such as stellar
gravitational collapse or collisions of compact stars and
black holes are needed to fully understand the incoming
wealth of observations from high-energy astronomy and
gravitational wave astronomy. It is thus not surprising
that in recent years hydrodynamical simulations of
compact objects in numerical relativity has become the
focus of several research
groups~\cite{Wilson96,Nakamura99a,Baumgarte99b,Font98b,Miller01a,Shibata99c,Shibata99d,Shibata99e}.
	
	In a previous paper~\cite{Font98b} (hereafter
paper I) we presented a 3D general-relativistic
hydrodynamics code (GR\_Astro) constructed for the NASA
Neutron Star Grand Challenge
Project~\cite{NASAreport}. The GR\_Astro code has been
developed by Washington University and the Albert
Einstein Institute and has the capability of solving the
coupled set of the Einstein equations and the general
relativistic hydrodynamic (GRHydro)
equations~\cite{GR3D}. It has been built using the Cactus
Computational Toolkit~\cite{Cactusweb} constructed by the
Albert Einstein Institute, Washington University and
other institutes.  Paper I presented our formulation for
the GRHydro equations coupled either to the standard
Arnowitt-Deser-Misner (ADM)~\cite{Arnowitt62} formulation
of the Einstein equations or to a hyperbolic formulation
of the equations~\cite{Bona97a}. It demonstrated the
consistency and convergence of the code for a
comprehensive sample of testbeds having analytic
solutions. It gave a detailed analysis of twelve
different combinations of spacetime and hydrodynamics
evolution methods, including Roe's and other approximate
Riemann solvers, as well as their relative performance
and comparisons when applied to the various testbeds. The
code as described and validated in paper I has been
applied to various physical problems, such as those
discussed in
Refs.~\cite{Miller01a,Alcubierre00b,Stergioulas01}, and
is now freely available~\cite{GR3D}.

	The main purpose of this paper is to examine and
validate our code in long-term, accurate simulations of
the dynamics of isolated stars in strong gravitational
fields. Single relativistic stars are indeed expected as
the end-point of a number of astrophysical scenarios
(such as gravitational collapse and binary neutron star
merging) and should provide important information about
strong field physics both through electromagnetic and
gravitational wave emissions. A number of new
numerical techniques have been incorporated in the
present code leading to a much improved
ability to simulate relativistic stars. These techniques
concern both the evolution of the field equations, for
which we have implemented new conformal-traceless
formulations of the Einstein equations, and in the
evolution of the hydrodynamical variables, for which the
use of the ``monotonized central differencing'' (MC)
limiter has provided us with the small error growth-rates
necessary for simulations over several dynamical
timescales.

	More precisely, in this paper we focus on the
accuracy of the code during long-term evolution
of spherical and rapidly rotating stellar
models.  We also investigate the nonlinear dynamics of
stellar models that are unstable to the fundamental
radial mode of pulsation. Upon perturbation, the unstable
models will either collapse to a black hole, or migrate
to a configuration in the stable branch of equilibrium
configurations (a behavior studied in the case of
unstable boson stars~\cite{Seidel90b}). In the case of
collapse, we follow the evolution all the way down to the
formation of a black hole, tracking the generation of its
apparent horizon. In the case of migration to the stable
branch, on the other hand, we are able to
accurately follow the nonlinear oscillations that accompany this
process and that can give rise to strong shocks. The
ability to simulate large amplitude oscillations is
important as we expect a neutron star formed in a
supernova core-collapse~\cite{Zwerger97,Dimmelmeier01} or
in the accretion-induced collapse of a white dwarf to
oscillate violently in its early stages of life.

	Particularly important for their astrophysical implications,
we study the linear pulsations of spherical and rapidly rotating
stars. The computed frequencies of radial, quasi-radial and
quadrupolar oscillations are compared with the corresponding
frequencies obtained with lower-dimensional numerical codes or with
alternative techniques such as the Cowling approximation (in which the
spacetime is held fixed and only the GRHydro equations are evolved) or
relativistic perturbative methods. The comparison shows an excellent
agreement confirming the ability of the code to extract physically
relevant information from tiny perturbations.  The successful determination
of the eigenfrequencies for rapidly rotating stars computed with our
code is noteworthy.  Such frequencies have not been obtained before
with the system being too complicated for perturbative techniques.

	The simulations discussed here make use of two
different polytropic equations of state (EOS). In
addition to the standard ``adiabatic'' EOS, in which the
pressure is expressed as a power law of the rest-mass
density, we have carried out simulations implementing the
``ideal fluid'' EOS, in which the pressure is
proportional to both the rest-mass density and the
specific internal energy density. This latter choice
increases the computational costs (there is one
additional equation to be solved) but allows for the
modeling of non-adiabatic processes, such as strong
shocks and the conversion of bulk kinetic energy into
internal energy, which are expected to accompany
relativistic astrophysical events.

	There are a number of reasons why we advocate the
careful validation of general relativistic astrophysics
codes.  Firstly, the space of solutions of the coupled
system of the Einstein and GRHydro equations is, to a large
extent, unknown.  Secondly, the numerical codes must
solve a complicated set of coupled partial differential
equations involving thousands of terms and there are
plenty of chances for coding errors. Thirdly, the complex
computational infrastructure needed for the use of the
code in a massively parallel environment, increases the
risk of computational errors, a risk that can only be
minimized through meticulous tests such as those
presented here as well as in paper I. This paper,
however, wants to be more than a list of testbeds: the
results presented show that our current numerical methods
are mature enough for obtaining answers to new and outstanding
problems in the physics of relativistic stars.

	The organization of this paper is as follows: the
formulation of the differential equations for the
spacetime and the hydrodynamics is briefly reviewed in
section~\ref{sec:equations}. Section~\ref{Numerical}
gives a short description of the numerical methods, with
emphases on the new schemes introduced in this paper (in
addition to those in paper I).
Sections~\ref{sec:simulations}--~\ref{gws} represent the
core of the paper and there we present the main results
of our simulations. In section~\ref{sec:simulations}, in
particular, we focus our attention on the simulation of
nonrotating relativistic stars. In section \ref{RRNSs} we
consider the evolution of rotating
stars. Section~\ref{gws} is dedicated to the extraction
of gravitational waveforms generated by the non-radial
pulsations of perturbed relativistic stars. In
section~\ref{sec:conclusion} we summarize our results and
conclusions. We use a spacelike signature
$(-,+,+,+)$ and units in which $c=G=M_\odot=1$ (geometric
units based on solar mass) unless explicitly
specified. Greek indices are taken to run from 0 to 3 and
Latin indices from 1 to 3.

\section{Basic Equations}
\label{sec:equations}                                                   %

We give a brief overview of the system of equations
in this section.  We refer the reader to paper I for 
more details.  

\subsection{Field Equations}
\label{feqs}

	In general relativity, the dynamics of the
spacetime is described by the Einstein field equations
$G_{\mu \nu} = 8 \pi T_{\mu \nu}$, with $G_{\mu \nu}$
being the Einstein tensor and $T_{\mu \nu}$ the
stress-energy tensor. Many different formulations of the
equations have been proposed throughout the years,
starting with the ADM formulation in
1962~\cite{Arnowitt62}. In our code, we have implemented
three different formulations of the field equations,
including the ADM formulation, a hyperbolic
formulation~\cite{Bona97a} and a more recent
conformal-traceless formulation based on the ADM
construction~\cite{Shibata95,Baumgarte99} (see also
Ref.~\cite{Alcubierre99d}).

	In the ADM formulation~\cite{Arnowitt62}, the
spacetime is foliated with a set of non-intersecting
spacelike hypersurfaces. Two kinematic variables relate
the surfaces: the lapse function $\alpha$, which
describes the rate of advance of time along a timelike
unit vector $n^\mu$ normal to a surface, and the shift
three-vector $\beta^i$ that relates the spatial
coordinates of two surfaces. In this construction the
line element reads
\begin{equation}
ds^2 = -(\alpha^{2} -\beta _{i}\beta ^{i}) dt^2 + 
        2 \beta_{i} dx^{i} dt +\gamma_{ij} dx^{i} dx^{j}
        \ .
\end{equation}
The original ADM formulation casts the Einstein equations
into a first-order (in time)
quasi-linear~\cite{Richtmyer67} system of equations. The
dependent variables are the 3-metric $\gamma_{ij}$ and
the extrinsic curvature $K_{ij}$. The evolution equations
read
\begin{eqnarray}
\partial_t \gamma_{ij} &=& - 2 \alpha K_{ij}+\nabla_i
        \beta_j + \nabla_j \beta_i, 
\label{dtgij} \\
        \partial_t K_{ij} &=& -\nabla_i \nabla_j \alpha + \alpha \Biggl[
        R_{ij}+K\ K_{ij} -2 K_{im} K^m_j  \nonumber \\
        &\ & - 8 \pi \left( S_{ij} - \frac{1}{2}\gamma_{ij}S \right)
        - 4 \pi {\rho}_{_{\rm ADM}} \gamma_{ij}
        \Biggr] \nonumber \\ 
        &\ & + \beta^m \nabla_m K_{ij}+K_{im} 
        \nabla_j \beta^m+K_{mj} \nabla_i \beta^m,
\label{dtkij}
\end{eqnarray}
where $\nabla_i$ denotes the covariant derivative with
respect to the 3-metric $\gamma_{ij}$, $R_{ij}$ is the
Ricci curvature of the 3-metric, and
$K\equiv\gamma^{ij}K_{ij}$ is the trace of the extrinsic
curvature. In addition to the evolution equations, there
are four constraint equations: the Hamiltonian constraint
\begin{equation}
\label{ham_constr}
{}^{(3)}R + K^2 - K_{ij} K^{ij} - 16 \pi
        {\rho}_{_{\rm ADM}} = 0,
\end{equation}
and the momentum constraints
\begin{equation}
\label{mom_constr}
\nabla_j K^{ij} - \gamma^{ij} \nabla_j K - 8 \pi j^i = 0.
\end{equation}
In equations (\ref{dtgij})--(\ref{mom_constr}),
${\rho}_{_{\rm ADM}},j^i,S_{ij},S \equiv \gamma^{ij}
S_{ij}$ are the components of the stress-energy tensor
projected onto the 3D hypersurface (for a more detailed
discussion, see Ref.~\cite{York79}).

	As mentioned above, in addition to the two
formulations described in paper I, we have recently
implemented a conformal-traceless reformulation of the
ADM system, as proposed by~\cite{Shibata95,Baumgarte99}.
Details of our particular implementation of this
formulation are extensively described in
Ref.~\cite{Alcubierre99d} and will not be repeated
here. We only mention here that this formulation makes
use of a conformal decomposition of the 3-metric,
\hbox{$\tilde \gamma_{ij} = e^{- 4 \phi} \gamma_{ij}$}
and the trace-free part of the extrinsic curvature,
\hbox{$A_{ij} = K_{ij} - \gamma_{ij} K/3$}, with the
conformal factor $\phi$ chosen to satisfy $e^{4 \phi} =
\gamma^{1/3} \equiv \det(\gamma_{ij})^{1/3}$. In this
formulation, as shown in Ref.~\cite{Baumgarte99}, in
addition to the evolution equations for the conformal
three--metric $\tilde \gamma_{ij}$ and the
conformal-traceless extrinsic curvature variables $\tilde
A_{ij}$, there are evolution equations for the conformal
factor $\phi$, the trace of the extrinsic curvature $K$
and the ``conformal connection functions'' $\tilde
\Gamma^i$ (following the notation of
Ref.~\cite{Baumgarte99}). We note that the final mixed,
first and second-order, evolution system for $\{\phi, K,
\tilde \gamma_{ij}, \tilde A_{ij}, \tilde \Gamma^i\}$ is
not in any immediate sense
hyperbolic~\cite{FriedrichPrivateComm}. In the original
formulation of Ref.~\cite{Shibata95}, the auxiliary
variables ${\tilde F}_i = -\sum_j \tilde \gamma_{ij,j}$
were used instead of the $\tilde \Gamma^i$.

        In Refs.~\cite{Alcubierre99d,Alcubierre99e} the
improved properties of this conformal-traceless
formulation of the Einstein equations were compared to
the ADM system. In particular, in
Ref.~\cite{Alcubierre99d} a number of strongly
gravitating systems were analyzed numerically with {\em
convergent} HRSC methods with {\it
total-variation-diminishing} (TVD) schemes using the
equations described in paper I. These included weak and
strong gravitational waves, black holes, boson stars and
relativistic stars. The results show that our treatment
leads to a stable numerical evolution of the many
strongly gravitating systems. However, we have also found
that the conformal-traceless formulation requires grid
resolutions higher than the ones needed in the ADM
formulation to achieve the same accuracy, when the
foliation is made using the ``K-driver'' approach
discussed in Ref.~\cite{Balakrishna96a}. Because in
long-term evolutions a small error growth-rate is the
most desirable property, we have adopted the
conformal-traceless formulation as our standard form for
the evolution of the field equations.

\subsection{Hydrodynamic Equations}
\label{heqs}

	The GRHydro equations are obtained from the local
conservation laws of the density current (continuity
equation) and of the stress-energy tensor, which we
assume to be that of a perfect fluid $T^{\mu\nu}= \rho h
u^{\mu}u^{\nu} + Pg^{\mu\nu}$, with $u^\mu$ being the
fluid 4-velocity and $P$ and $h$ the (isotropic) pressure
and the specific enthalpy, respectively. In our code the
GRHydro equations are written as a first-order
flux-conservative hyperbolic
system~\cite{Banyuls97,Font98b}
\begin{equation}
{\partial}_t \vec{\cal{U}} + {\partial}_i \vec{F^i} =
\vec{S} \ ,
\label{balance}
\end{equation}
where the evolved state vector $\vec{\cal{U}}$ is given,
in terms of the {\it primitive} variables: the rest-mass
density $\rho$, the 3-velocity $v^i = u^i/W
+\beta^i/\alpha$ and the specific internal energy
$\varepsilon$, as
\begin{equation}
\label{eq:evolvedvar}
\vec{\cal{U}} = 
\left[ 
\begin{array}{c}
        \tilde{D} \\ \\ \tilde{S_j} \\ \\ \tilde{\tau} \\
        \\ \end{array} \right]= \left[ \begin{array}{c}
        \sqrt{\gamma} W \rho \\ \\ \sqrt{\gamma} \rho h
        W^2 v_j \\ \\ \sqrt{\gamma} (\rho h W^2 - P - W
        \rho) \\ \\ \end{array} \right] \ .
\end{equation}
Here $\gamma$ is the determinant of the 3-metric
$\gamma_{ij}$ and $W$ is the Lorentz factor, $W=\alpha
{u^0} = {(1 - \gamma_{ij} v^i v^j)}^{-1/2}$. Furthermore, the 3-flux
vectors $\vec{F^i}$ are given by
\begin{equation}
\label{hydroflux}
        \vec{F^i} = \left[ \begin{array}{c}
        \alpha (v^i - \frac {1}{\alpha} {\beta}^i) \tilde{D}  \\ \\
        \alpha [(v^i - \frac {1}{\alpha} {\beta}^i) 
        \tilde{S_j} + \sqrt{\gamma} P {\delta}^i_j]            \\ \\
        \alpha [(v^i - \frac {1}{\alpha} {\beta}^i) 
        \tilde{\tau} + \sqrt{\gamma} v^i P]
\end{array}  \right].
\end{equation}
Finally, the source vector $\vec{S}$ is given by
\begin{equation}
\vec{S} = \left[ \begin{array}{c}
        0                                               \\ \\
        \alpha \sqrt{\gamma} T^{\mu \nu} g_{\nu \sigma}
        { {\Gamma}^{\sigma} }_{\mu j}                       \\ \\
        \alpha \sqrt{\gamma} (T^{\mu 0} {\partial}_{\mu} \alpha -
        \alpha T^{\mu \nu} { {\Gamma}^0}_{\mu \nu})
\end{array}  \right],
\end{equation}
where ${ {\Gamma}^{\alpha} }_{\mu \nu}$ are the
Christoffel symbols.

\subsection{Gauge Conditions}

	The code is designed to handle arbitrary shift
and lapse conditions, which can be chosen as appropriate
for a given spacetime simulation. More information about
the possible families of spacetime slicings which have
been tested and used with the present code can be found
in Refs.~\cite{Font98b,Alcubierre99d}. Here, we limit
ourselves to recall details about the specific foliations
used in the present evolutions. In particular, we have
used algebraic slicing conditions of the form
\begin{equation}
(\partial_t - \beta^i\partial_i) \alpha = - f(\alpha) \;
\alpha^2 K,
\label{eq:BMslicing}
\end{equation}
with $f(\alpha)>0$ but otherwise arbitrary. This choice
contains many well known slicing conditions.  For
example, setting $f=1$ we recover the ``harmonic''
slicing condition, or by setting \mbox{$f=q/\alpha$},
with $q$ being an integer, we recover the generalized
``1+log'' slicing condition~\cite{Bona94b} which after
integration becomes
\begin{equation}
\label{1+log}
\alpha = g(x^i) + \frac{q}{2}\log \gamma, 
\end{equation}
where $g(x^i)$ is an arbitrary function of space only. In
particular, all of the simulations discussed in this
paper are done using condition (\ref{1+log}) with $q=2$,
basically due to its computational efficiency (we caution
that ``gauge pathologies'' could develop with the
``1+log'' slicings, see
Ref.~\cite{Alcubierre97a,Alcubierre97b}).

The evolutions presented in this paper were carried out
with the shift vector being either zero or constant in time.

\section{Numerical methods}
\label{Numerical}

        We now briefly describe the numerical schemes
used in our code. We will distinguish the schemes
implemented in the evolution of the Einstein equations
from those implemented in the evolution of the
hydrodynamic equations. In both cases, the equations are
finite-differenced on spacelike hypersurfaces covered
with 3D numerical grids using Cartesian coordinates.

\subsection{Spacetime Evolution}

	As described in paper I, our code supports the
use of several different numerical
schemes~\cite{Font98b,Alcubierre99d}.  Currently, a
Leapfrog (non-staggered in time) and an iterative
Crank-Nicholson scheme have been coupled to the
hydrodynamic solver.

	The Leapfrog method assumes that all variables
exist on both the current time step $t^n$ and the
previous time step $t^{n-1}$. Variables are updated from
$t^{n-1}$ to $t^{n+1}$ (future time) evaluating all terms
in the evolution equations on the current time step
$t^n$. The iterative Crank-Nicholson solver, on the other
hand, first evolves the data from the current time step
$t^n$ to the future time step $t^{n+1}$ using a forward
in time, centered in space first-order method.  The
solution at steps $t^n$ and $t^{n+1}$ are then averaged
to obtain the solution on the half time step
$t^{n+1/2}$. This solution at the half time step
$t^{n+1/2}$ is then used in a Leapfrog step to re-update
the solution at the final time step $t^{n+1}$. This
process is then iterated. The error is defined as the
difference between the current and previous solutions on
the half time step $t^{n+1/2}$. This error is summed over
all gridpoints and all evolved variables. Because the
smallest number of iterations for which the iterative
Crank-Nicholson evolution scheme is stable is three and
further iterations do not improve the order of
convergence~\cite{Teukolsky00,Alcubierre99d}, we do not
iterate more than three times. Unless otherwise noted, 
all simulations reported in this paper use the
iterative Crank-Nicholson scheme for the time evolution
of the spacetime.

\subsection{Hydrodynamical Evolution}

        The numerical integration of the GRHydro equations is
based on High-Resolution Shock-Capturing (HRSC) schemes,
specifically designed to solve nonlinear hyperbolic
systems of conservation laws. These conservative schemes
rely on the characteristic structure of the equations in
order to build approximate Riemann solvers. In paper I we
presented a spectral decomposition of the GRHydro equations,
suitable for a general spacetime metric (see also
Ref.~\cite{Ibanez01}).

	Approximate Riemann solvers compute, at every
cell-interface of the numerical grid, the solution of
local Riemann problems (i.e. the simplest initial value
problem with discontinuous initial data). Hence HRSC
schemes automatically guarantee that physical
discontinuities developing in the solution (e.g., shock
waves, which appear in core-collapse supernovae or in
coalescing neutron star binaries) are treated
consistently. HRSC schemes surpass traditional approaches
\cite{Wilson96,Shibata99c} which rely on the use of
artificial viscosity to resolve such discontinuities,
especially for large Lorentz factor flows. HRSC schemes
have a high order of accuracy, typically second-order or
more, except at shocks and extremal points.  We refer the
reader to~\cite{Ibanez99,Font00} for recent reviews on
the use HRSC schemes in relativistic hydrodynamics.

	One of the major advantages of using HRSC schemes
is that we can take advantage of the many different
algorithms that have been developed and tested in
Newtonian hydrodynamics. In this spirit, our code allows
for three alternative ways of performing the numerical
integration of the hydrodynamic equations: {\it (i)}
using a flux-split method~\cite{Hirsch92}; {\it (ii)}
using Roe's approximate Riemann solver~\cite{Roe81}, and
{\it (iii)} using Marquina's flux-formula~\cite{Donat96}.
The different methods differ simply in the way the
numerical fluxes at the cell-interfaces are calculated in
the corresponding flux-formula. The code uses
slope-limiter methods to construct second-order TVD
schemes~\cite{Harten84} by means of monotonic piecewise
linear reconstructions of the cell-centered quantities to
the left (L) and right (R) sides of every cell-interface
for the computation of the numerical fluxes. More
precisely, $\vec{\cal{U}}_i^R$ and
$\vec{\cal{U}}_{i+1}^L$ are computed to second-order
accuracy as follows:
\begin{eqnarray}
\vec{\cal{U}}^R_i &=& \vec{\cal{U}}_i + \sigma_i (x_{i+\frac{1}{2}}-x_i)
\\
\vec{\cal{U}}^L_{i+1} &=& \vec{\cal{U}}_{i+1} +
\sigma_{i+1} (x_{i+\frac{1}{2}}-x_{i+1})
\end{eqnarray}
\noindent
where $x$ denotes a generic spatial coordinate. We have
focused our attention on two different types of slope
limiters, the standard ``minmod'' limiter and the
``monotonized central-difference" (MC)
limiter~\cite{vanLeer77}. In the first case, the slope
$\sigma_i$ is computed according to
\begin{equation}
\sigma_i = \mbox{minmod}\left(\frac{\vec{\cal{U}}_i-
\vec{\cal{U}}_{i-1}}{\Delta x},
\frac{\vec{\cal{U}}_{i+1}-\vec{\cal{U}}_{i}}{\Delta x}\right),
\end{equation}
\noindent
where $\Delta x$ denotes the cell spacing.
The minmod function of two arguments is defined by
\[
\mbox{minmod}(a,b) \equiv \left\{ \begin{array}{cl} 
        a & \mbox{if $|a|<|b|$ and $ab>0$} \\ \\ 
        b & \mbox{if $|b|<|a|$ and $ab>0$} \\ \\ 
        0 & \mbox{if $ab\leq 0$}
     \end{array} \right.
\]

        On the other hand, the MC slope limiter (which
was not included in the previous version of the code
discussed in paper I) does not reduce the slope as
severely as minmod near a discontinuity and, therefore, a
sharper resolution can be obtained. In this case the
slope is computed as
\begin{equation}
\sigma_i = \mbox{MC}\left(\frac{\vec{\cal{U}}_i-
\vec{\cal{U}}_{i-1}}{\Delta x},
\frac{\vec{\cal{U}}_{i+1}-\vec{\cal{U}}_{i}}{\Delta x}\right),
\end{equation}
where the MC function of two arguments is defined by
\[
\mbox{MC}(a,b) \equiv \left\{ \begin{array}{cl} 
        2a & \mbox{if $|a|<|b|$,\ \ and $2|a|<|c|$, and $ab>0$} \\ \\ 
        2b & \mbox{if $|b|<|a|$,\ \ and $2|b|<|c|$, and $ab>0$} \\ \\ 
        c  & \mbox{if $|c|<2|a|$,   and $|c|<2|b|$, and $ab>0$} \\ \\ 
        0  & \mbox{if $ab\leq 0$}
     \end{array} \right.
\]
and where $c\equiv (a+b)/2$. Both schemes provide the
desired second-order accuracy for smooth solutions, while
still satisfying the TVD property. In sect.~\ref{sphstar}
we will report on a comparison between the two algorithms
and justify the use of the MC slope limiter as our
preferred one.

\subsection{Equations of State}
\label{eos}

	As mentioned in the Introduction, to explore the
behavior of our code in long-term evolutions of
equilibrium configurations, we used two different
polytropic equations of state and at various central
rest-mass densities. In particular, we have implemented
both an {\it adiabatic} (or zero temperature) EOS
\begin{equation}
\label{adiabatic}
P=K\rho^{\Gamma} = K\rho^{1 + 1/N}\ , 
\end{equation}
and as a so-called {\it ``ideal fluid''} EOS
\begin{equation}
\label{idealf}
P=(\Gamma-1)\rho\varepsilon \ , 
\end{equation}
where $K$ is the polytropic constant, $\Gamma$ the
polytropic index and $N\equiv(\Gamma-1)^{-1}$ the
polytropic exponent. The ideal fluid EOS (\ref{idealf})
depends on both the rest-mass density $\rho$ and on the
specific internal energy $\varepsilon$, it corresponds to 
allowing the polytropic coefficient $K$
in adiabatic EOS (\ref{adiabatic}) to be a function of entropy.
The use of an adiabatic EOS with a constant $K$ is 
computationally less expensive and
is physically reasonable when modeling configurations
that are in near equilibrium, such as stable stellar
models in quasi-equilibrium evolutions. 
There are however dynamical processes, such as
those involving nonlinear oscillations and shocks, in
which the variations in the energy entropy cannot be
neglected. The simulations discussed in
section~\ref{migration}, where both equations of state
(\ref{adiabatic})-(\ref{idealf}) are used for the same
configuration, gives direct evidence of how a more
realistic treatment of the internal energy of the system
can produce qualitatively different results.

	The increased accuracy in the physical
description of the dynamical system comes with a
non-negligible additional computational cost. It
involves the solution of an additional equation
(i.e. the evolution equation for the specific internal
energy $\varepsilon$), increasing the total number of GRHydro
equations from four to five and making accurate long-term
evolutions considerably harder.

\subsection{Boundary Conditions}
\label{bcs}

	In our general-purpose code, a number of different
boundary conditions can be imposed for either the
spacetime variables or for the hydrodynamical variables.
We refer the reader to ~\cite{Font98b,Alcubierre99d} for
details. In all of the runs presented in this paper we
have used static boundary conditions for the
hydrodynamical variables and radiative outgoing boundary
conditions for the spacetime variables. The only
exception to this is the evolution of rotating stars (see
sect.~\ref{RRNSs}), for which the spacetime variables
have also been held fixed at the outer boundary.

\section{Spherical Relativistic Stars}               %
\label{sec:simulations}                              %

We turn next to the description of the numerical
evolutions of relativistic star configurations. We 
start by considering spherical models.

\subsection{Long-term evolution of stable configurations}
\label{sphstar}

        Using isotropic coordinates $(t,r,\theta,\phi)$,
the metric describing a static, spherically symmetric
relativistic star reads
\begin{equation} 
\label{isot}
ds^2 =  -e^{2 \nu} dt^2 + e^{2 \lambda }(dr^2+r^2 d \theta^2 
        +r^2 \sin^2 \theta  d \phi^2), 
\end{equation}
where $\nu$ and $\lambda$ are functions of the radial
coordinate $r$ only. The form of the metric component
$g_{rr}$ is much simpler in these coordinates than in
Schwarzschild coordinates, which are often used to
describe a Tolman-Oppenheimer-Volkoff (TOV) equilibrium
stellar solution. In addition, $g_{rr}$ is not
constrained to be equal to unity at the center of the
stellar configuration, as in Schwarzschild
coordinates. We have found that these two properties of
the isotropic coordinates are very beneficial to achieve
long-term numerical evolutions of relativistic
stars. Therefore, all simulations of spherical
relativistic stars shown in this paper have been
performed adopting the line element (\ref{isot})
expressed in Cartesian coordinates.

\begin{figure}[htb]
\centerline{\psfig{file=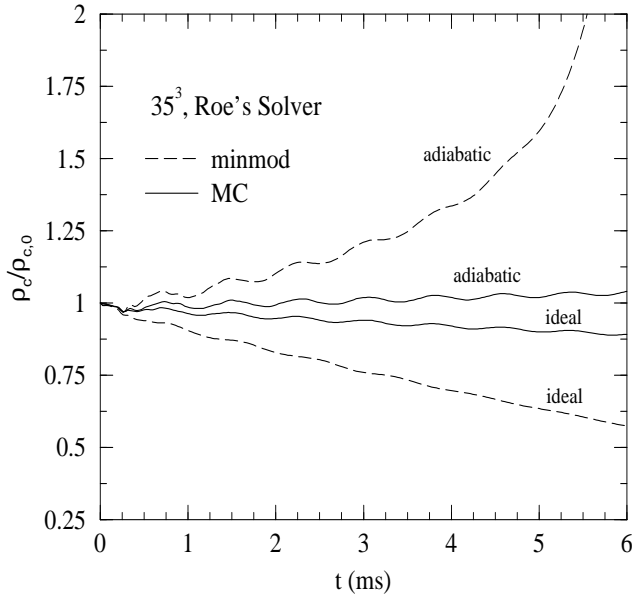,height=3.2in,width=3.3in}}
\caption{Evolution of the central rest-mass density
$\rho_c$ (in units of the initial central rest-mass
density $\rho_{c,0}$) for a nonrotating star with
gravitational mass $M=1.65\;M_\odot$. Using Roe's
approximate Riemann solver, the figure shows a comparison
in the use of the minmod and of the MC slope limiters for
both the ideal fluid and the adiabatic EOS. }
\label{fig:rho_eos_cf}
\end{figure}

	Although the initial configurations refer to
stellar models in stable equilibrium, the truncation
errors at the center and at the surface of the star
excite small radial pulsations that are damped in time by
the numerical viscosity of the code. Moreover, these
pulsations are accompanied by a secular evolution of the
values of the central rest-mass density away from its
initial value. Similar features have been reported in
Refs.~\cite{Stergioulas99,Font99}.  These features
converge away at
the correct rate with increasing grid resolution and do
not influence the long-term evolutions. Moreover, the
secular evolution of the central rest-mass density varies
according to the EOS adopted: when using the ideal fluid
EOS, we have observed that the secular drift of the
central rest-mass density is towards lower
densities. However, if we enforced the adiabatic
condition (which is justified for the case of a
near-equilibrium evolution), we have observed that the
dominant truncation error has opposite sign and the
central rest-mass density evolves towards larger values.
 
\begin{figure}[htb]
\centerline{\psfig{file=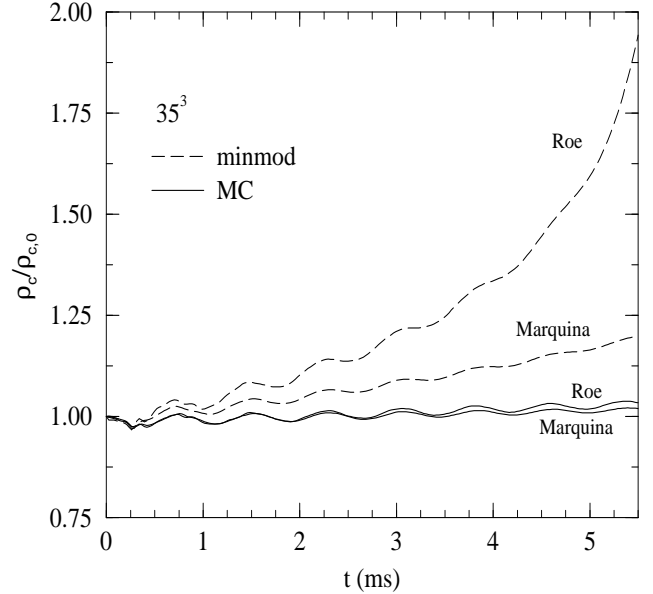,height=3.2in,width=3.3in}}
\caption{Evolution of the normalized central rest-mass
density $\rho_c$ for a nonrotating $M=1.65\;M_\odot$
star. Different lines show a comparison between Roe's
Riemann solver and Marquina's flux-formula for different
slope limiters.}
\label{fig:rho_marquina_roe_cf}
\end{figure}
        
        This is shown in Fig.~\ref{fig:rho_eos_cf} where
we plot the evolution of a TOV star with gravitational
mass $M=1.65\;M_{\odot}$, constructed with a $N=1$
polytrope. In our units, the polytropic constant is
$K=123.5$ and the central rest-mass density of the star
is $\rho_c=1.00 \times 10^{-3}$. For these tests, a very
coarse grid of $35^3$ gridpoints in octant symmetry is
sufficient and allows the major effects to be revealed
with minimal computational costs. The outer boundary is
placed at about $1.7\;r_s$ (where $r_s$ is the isotropic
coordinate radius of the star). We use radiative boundary
conditions with a $1/r$ fall-off. Irrespective of the
slope limiter used, the magnitude of the secular drift
observed in the central rest-mass density evolution is
roughly a factor of two smaller for the adiabatic EOS
than for the ideal fluid EOS. As a result, in all of the
evolutions of stable configurations which remain close to
equilibrium (such as pulsating stars, with no shock
developing), the adiabatic EOS is preferred.

        Fig.~\ref{fig:rho_eos_cf} also gives a comparison
of the use of the minmod and the MC slope limiters in the
evolution of the normalized central rest-mass
density. For both the ideal fluid and the adiabatic EOS,
the MC limiter shows a significantly smaller secular
increase in the central rest-mass density, as compared to
the minmod one. The simulations in
Fig.~\ref{fig:rho_eos_cf} employed Roe's approximate
Riemann solver in the fluid evolution scheme and this is
then compared to Marquina's flux-formula in
Fig.~\ref{fig:rho_marquina_roe_cf} for the evolution of
the central rest-mass density. The secular increase is
significantly smaller when using Marquina's flux-formula
than when using Roe's solver, and this is especially
noticeable for the minmod slope limiter. A comparison of
the increase of the maximum error in the Hamiltonian
constraint after several ms of evolution (not shown here)
indicates that it is about $80 \% $ smaller with Marquina
than with Roe, when using the adiabatic EOS. As a result
of the above comparisons, we have adopted Marquina's
scheme with the MC slope limiter as our preferred scheme
for evolution of the GRHydro equations. Unless otherwise
noted, all of the simulations presented in this paper
have been obtained with such a scheme.

\begin{figure}[htb]
\centerline{\psfig{file=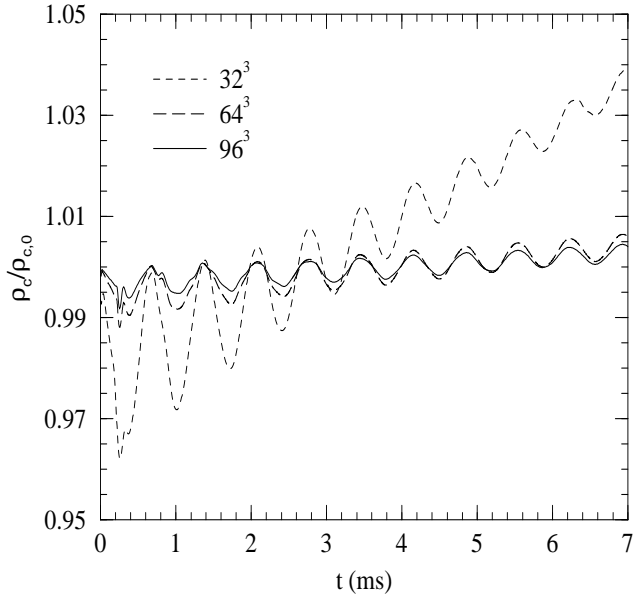,height=3.2in,width=3.3in}}
\caption{Time evolution of the normalized central
rest-mass density at three different grid resolutions
($32^3$, $64^3$ and $96^3$ gridpoints, respectively), for
a $M=1.4\;M_{\odot}$, $N=1$ relativistic, spherical
polytrope. The evolution of the central rest-mass density
is mainly modulated by the fundamental radial mode of
oscillation of the star. The initial amplitude of the
oscillation converges to zero at second-order, while the
secular increase in the central rest-mass density
converges away at almost second-order.}
\label{fig:rho_maxN1.0.eps}
\end{figure}

        Next, we show in Fig.~\ref{fig:rho_maxN1.0.eps}
the long-term evolution of the central rest-mass density
for three different grid resolutions. For this, we
consider a nonrotating $N=1$ polytropic star with
gravitational mass $M=1.4\;M_\odot$, circumferential
radius $R=14.15$ km, central rest-mass density
$\rho_c=1.28 \times 10^{-3}$ and $K=100$. The different
simulations used $32^3$, $64^3$ and $96^3$
gridpoints with octant symmetry and with the outer
boundary placed at $1.7\;r_s$. These grid resolutions
correspond to about 19, 38 and 56 gridpoints per star
radius, respectively.  Fig.~\ref{fig:rho_maxN1.0.eps}
shows the oscillations in the central rest-mass density
and the secular evolution away from the initial value
mentioned above. The oscillations are produced by the
first-order truncation error at the center and the
surface of the star (our hydrodynamical evolution schemes
are globally second order, but only first-order at local
extrema; see related discussions in
Ref.~\cite{Alcubierre99d}, where long-term convergence
tests are presented) but both the amplitude of the
initial oscillation and the rate of the secular change
converge to zero at nearly second-order with increasing
grid resolution.

\begin{figure}[htb]
\centerline{\psfig{file=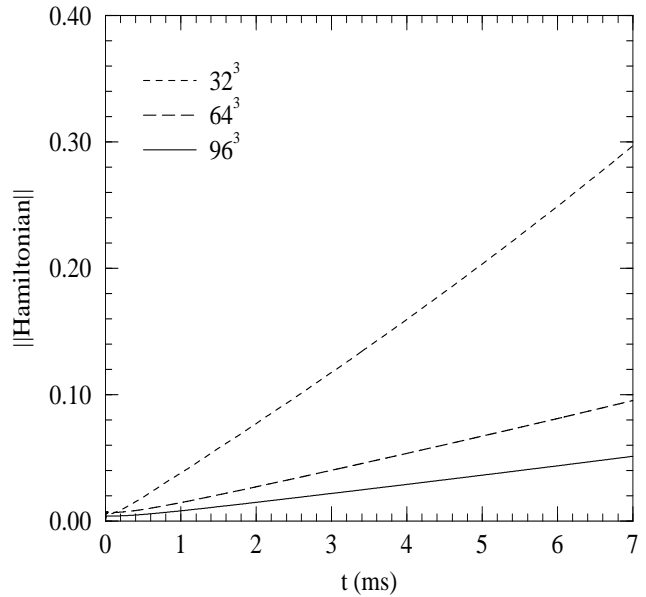,height=3.2in,width=3.3in}}
\caption{Convergence of the L2-norm of the Hamiltonian
constraint, at three different grid resolutions ($32^3$,
$64^3$ and $96^3$ gridpoints, respectively), for a
$M=1.4\;M_{\odot}$, $N=1$ polytropic spherical
relativistic star. The rate of convergence is close to
second-order with increasing grid resolution.}
\label{fig:ham_nm2N1.0}
\end{figure}

\begin{figure}[htb]
\centerline{\psfig{file=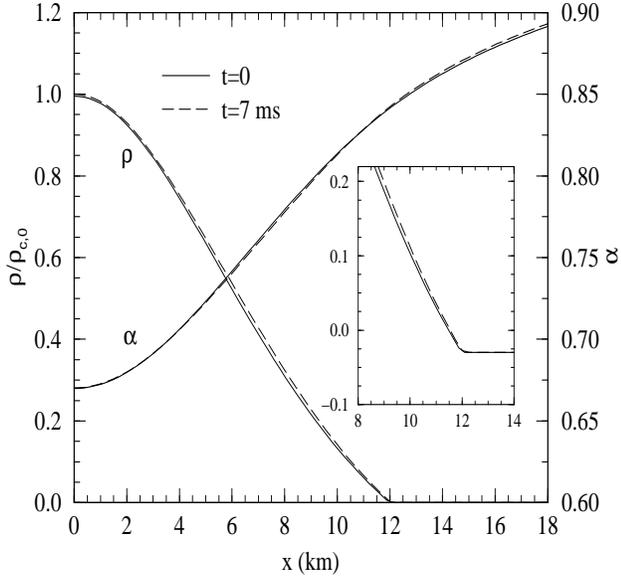,height=3.2in,width=3.3in}}
\caption{ Variation of the original profiles along the
$x$-axis of the rest-mass density (left vertical axis)
and lapse function (right vertical axis), for a
$M=1.4\;M_{\odot}$, $N=1$ polytropic spherical
relativistic star, after 7 ms of evolution. A $96^3$ grid
in octant symmetry was used in the simulation. The small
inset shows a magnification of the rapid change in the
gradient of the rest-mass density profile at the surface
of the star. }
\label{fig:rho_alp_t0_7msN1.0.eps}
\end{figure}

	Note that the evolutions shown in
Figs.~\ref{fig:rho_maxN1.0.eps}-\ref{fig:rho_alp_t0_7msN1.0.eps}
extend to 7 ms, corresponding to about 10 dynamical times
(taking the fundamental radial mode period of pulsation
as a measure of the dynamical timescale), significantly
longer than, say, the ones reported by other
authors~\cite{Shibata99c,Baumgarte99c}. Our evolutions
are limited by the time available (a simulation with
$96^3$ gridpoints and up to 7 ms takes about 40 hours on
a 128 processor Cray-T3E supercomputer.).  We have found
that for a resolution of $96^3$, the central density at
the end of the 7ms evolution is just 0.25\% larger than
the initial central density.

	For the same configuration, we show, in
Fig.~\ref{fig:ham_nm2N1.0}, the time evolution of the
L2-norm of the violation of the Hamiltonian constraint at the
three different grid resolutions. Also in this case, the
violation of the Hamiltonian constraint converges to zero
at nearly second-order with increasing grid resolution.

In Fig.~\ref{fig:rho_alp_t0_7msN1.0.eps}, we show other
aspects of the accuracy of the simulation with $96^3$
gridpoints, by comparing the initial profiles of the
rest-mass density $\rho$ and of the lapse function
$\alpha$ of the TOV star with those obtained after 7 ms
of evolution. The small deviations from the original
profiles are worth emphasizing. The small inset shows a
magnification of the rapid change in the gradient of the
rest-mass density profile at the surface of the star.

\subsection{Radial pulsations}
\label{rp}

        As mentioned in the previous section, the
truncation errors of the hydrodynamical schemes used in
our code trigger radial pulsations of the initially
static relativistic star (see Ref.~\cite{Kokkotas99} for
a review). These pulsations are initiated at the surface
of the star, where the gradients of the rest-mass density
are the largest
(cf. Fig.~\ref{fig:rho_alp_t0_7msN1.0.eps}). Because
gravitational waves cannot be emitted through the
excitation of radial pulsations of nonrotating
relativistic stars, these pulsations are damped only by
the numerical viscosity of the code in numerical
simulations of inviscid stars. In treatments more
dissipative than the HRSC schemes used in our code, such
as those using artificial viscosity or particle methods
(e.g. Smoothed Particle Hydrodynamics), these
oscillations will be damped significantly faster.

\begin{figure}[htb]
\centerline{\psfig{file=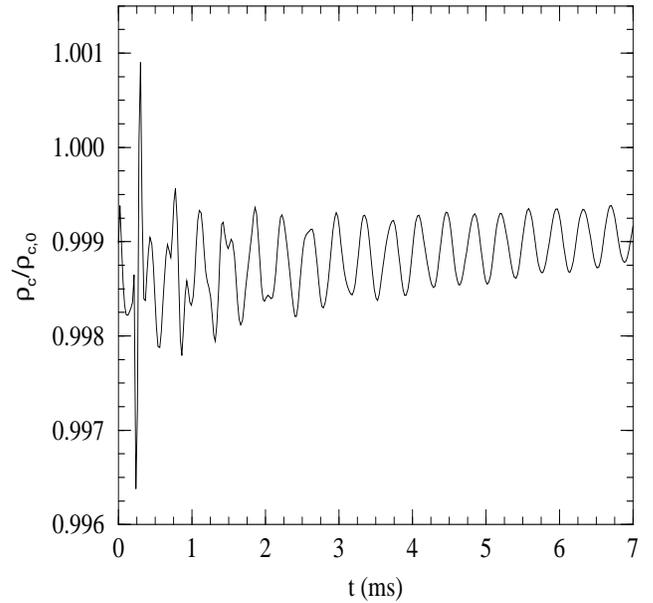,height=3.2in,width=3.3in}}
\caption{ Time evolution of the central rest-mass density
of a $M=1.4\;M_{\odot}$, $N=1$ polytropic spherical
relativistic star. In this the simulation the spacetime
is held {\it fixed} and the hydrodynamic variables have
been evolved on a numerical grid of $96^3$
gridpoints. The evolution is a superposition of radial
normal modes of pulsation, excited by truncation errors
of the hydrodynamical scheme. Higher overtones are damped
faster by the small but non-zero numerical viscosity.}
\label{fig:rho_max_fsN1.0.eps}
\end{figure}

        In order to test the properties of the long-term
hydrodynamical evolution separately from those of the
spacetime evolution, we have first examined the long-term
hydrodynamical evolution separately from those of the
spacetime evolution, we have first examined the
small-amplitude radial pulsations in a {\it fixed
spacetime} of an initially static relativistic star. As
initial data, we use the $M=1.4\;M_\odot$ polytropic
star of the previous section. We show, in
Fig.~\ref{fig:rho_max_fsN1.0.eps}, the evolution up to 7
ms of the normalized star's central rest-mass density
with a numerical grid of $96^3$ gridpoints. The amplitude of
the excited pulsations in this purely hydrodynamical
evolution is minute (less than 1 part in 200) and is
significantly smaller than the corresponding amplitude in
a coupled hydrodynamical and spacetime evolution (compare
the vertical axes of Figs.~\ref{fig:rho_maxN1.0.eps}
and~\ref{fig:rho_max_fsN1.0.eps}). 

        A closer look at Figure
\ref{fig:rho_max_fsN1.0.eps} reveals that the evolution
of the central rest-mass density is a superposition of
different radial normal modes of pulsation. The
higher-frequency modes are damped faster, so that after a
certain time the evolution proceeds mainly in the
fundamental mode of pulsation. Note also the small
damping rate of the fundamental pulsation mode,
indicating the small effective numerical viscosity of our
HRSC hydrodynamical scheme. The evolution towards larger
values of the central rest-mass density is similar to
that discussed in Section~\ref{sphstar} but less
pronounced in this case. At a resolution of $96^3$
gridpoints, the secular change in the average central
rest-mass density is less than 0.02\% for the total
evolution time shown.

	The use of truncation error as an initial
perturbation deserves commenting on. The oscillations
caused by truncation error will converge away with
increasing resolution, hence the overall oscillation
amplitude can carry no physical information about the
system. However, the frequencies and normalized
eigenfuntions of particular normal-modes of oscillation
of the star are physical (in the sense that they match
the eigenfrequencies and eigenfunctions calculated
through perturbative analyses) and can be extracted from
these simulations by carrying out a Fourier transform of
the time evolution of the radial velocity or of the
rest-mass density. As the small-amplitude pulsations are
in the linear regime, the eigenfunctions can be
normalized arbitrarily (e.g. to 1.0 at the surface of the
star). At increasing resolution, the solution converges
to the mode-frequencies and to the normalized
eigenfunctions, even though the overall oscillation
amplitude converges to zero. Such evolutions are useful
for extracting the properties of linear normal-modes of
oscillation, as long as the resolution is fine enough
that the pulsations excited by truncation errors are in
the linear regime and as long as the resolution is coarse
enough that the various local 1st and 2nd order
truncation errors of the numerical scheme result in a
time evolution that is dominated by a sum of normal modes
(at very fine resolutions the Fourier transform of the
time evolution would be very small and thus have a very
noisy power spectrum due to roundoff errors, in which
case the physical normal-mode frequencies would be
difficult to extract - this has not been the case for the
resolutions used in this paper). We also note that
different variants of our hydrodynamical evolution
schemes excite the various physical normal-modes at
different amplitudes. For example, 2nd order schemes
employing the minmod limiter tend to clearly excite a
large number of high-frequency overtones, whereas the use
of the MC limiter results in the clear excitation of only
a few low-frequency overtones and a more noisy FFT power
spectrum at higher frequencies (for the resolutions used
in this paper). This difference in behaviour is due to
the differences in the local truncation errors inherent
in these numerical schemes.

The radial pulsations are a sum of eigen modes of
pulsation. Since the radial pulsations triggered by truncation errors
have a small amplitude, one can compare the frequencies with that
computed by linear perturbation theory~\cite{Font99} or with
hydrodynamical evolutions of similar models in
2D~\cite{Stergioulas99,Font99}. In this way we can validate that the
``artificial'' perturbations produced by the truncation errors do
excite ``physical'' modes of oscillation for a relativistic
star. However, before discussing the results of this comparison, it is
important to emphasize that the identification of the frequency peaks
in the Fourier transform of the time evolution of a given variable
with physical frequencies must be done with care. A real pulsation
frequency must be global (the same at every point in the star, at
least for discrete normal mode frequencies) and it should appear in
the time evolution of different physical quantities describing the
star's structure and dynamics. To eliminate possible ambiguities, we
have carried out our frequency identification procedure for different
variables and at different positions in the star.

\begin{figure}[htb]
\centerline{\psfig{file=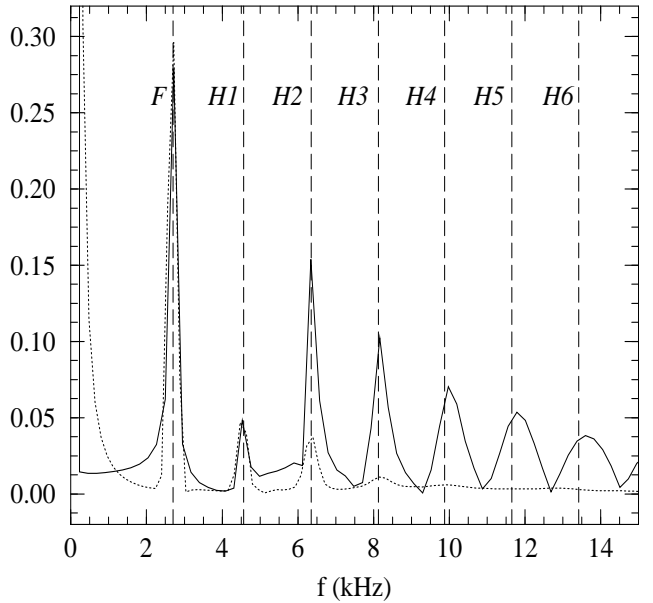,height=3.2in,width=3.3in}}
\caption{Fourier transform of the central rest-mass
density evolution of a $M=1.4\;M_{\odot}$, $N=1$
polytropic spherical relativistic star, in a {\it fixed
spacetime} evolution. Here $F$ represents the fundamental
normal mode frequency, while $H1-H6$ indicate the first
six overtones. The frequency peaks in the power spectrum
are in excellent agreement with the radial normal mode
frequencies (shown here as dashed vertical lines)
computed with an independent 2D code using spherical
polar coordinates. The solid and dotted lines were
computed with $96^3$ and $64^3$ gridpoints,
respectively. The units of the vertical axis are
arbitrary.}
\label{fig:rho_fft_fsN1.0.eps}
\end{figure}

	Fig.~\ref{fig:rho_fft_fsN1.0.eps} shows the
Fourier transform of the time evolution of the central
rest-mass density of the same initial model as in
Fig.~\ref{fig:rho_max_fsN1.0.eps}, but using the {\it
minmod} limiter (which gives a clearer excitation of the
higher overtones). We indicate with $F$ the fundamental
normal mode frequency and with $H1-H6$ the next six
higher frequency modes (overtones). We have also compared
the frequency peaks in the Fourier spectrum to both the
normal mode frequencies expected by linear perturbation
theory in the Cowling approximation~(see
Ref.~\cite{McDermott}), and to the frequencies computed
with an independent 2D axisymmetric nonlinear
code~\cite{Font99}, which uses the same HRSC schemes but
in spherical polar coordinates (shown as dashed vertical
lines in Fig.~\ref{fig:rho_fft_fsN1.0.eps}).

As can be seen from Table~\ref{table:radial}, the
agreement is extremely good. The relative difference
between the 3D and 2D results at this grid resolution is
better than $1\%$ up to ($H4$) and slightly larger for
higher frequencies ($H5$ and $H6$), which become
under-resolved at this grid resolution. This excellent
agreement is a significant test for the correct
implementation of the hydrodynamical evolution schemes in
our code, and is an indication of the level of accuracy
we can achieve, resolving and following these small
deviations away from the equilibrium configuration.  As
one would expect, lower or higher resolution runs
(e.g. with $64^3$ or $144^3$ gridpoints), which have
intrinsically larger or smaller perturbation amplitudes,
respectively, reproduce the peaks in the power spectrum
shown in Fig.~\ref{fig:rho_fft_fsN1.0.eps} (see dotted
line in Fig.~\ref{fig:rho_fft_fsN1.0.eps}, which
corresponds to an evolution with $64^3$ grid-points.
\begin{figure}[htb]
\centerline{\psfig{file=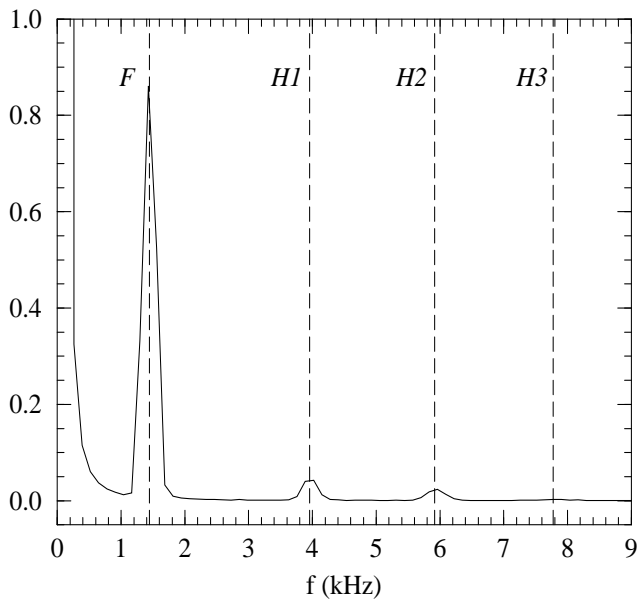,height=3.2in,width=3.3in}}
\caption{Fourier transform of the evolution of the radial
velocity for a $M=1.4\;M_{\odot}$, $N=1$ polytropic
spherical relativistic star in a {\it coupled} spacetime
and hydrodynamical evolution. The frequency peaks in the
spectrum are in excellent agreement with the radial
normal mode frequencies computed by perturbation theory
(shown here as dashed vertical lines). As in
Fig.~\ref{fig:rho_fft_fsN1.0.eps}, here $F$ represents
the fundamental normal mode frequency, while $H1-H3$ are
the next three higher frequency modes. The units of the
vertical axis are arbitrary.}
\label{fig:rho_fftN1.0.eps}
\end{figure}

\begin{table}[htb]
\caption{ Comparison of small-amplitude radial pulsation
frequencies obtained with the present 3D nonlinear
evolution code with frequencies obtained with an
independent 2D code. Both codes evolve the GRHydro equations
in a {\it fixed spacetime} and for an equilibrium model
of a $N=1$ relativistic polytrope with $M/R=0.15$. }
\medskip
\begin{tabular}{cccc}
Mode & Present 3D code & 2D code & Relative Difference \\
     & (kHz) & (kHz) & (\%)              \\ 
\hline 
        $F$  & 2.696  & 2.701  & 0.2      \\ 
        $H1$ & 4.534  & 4.563  & 0.6      \\ 
        $H2$ & 6.346  & 6.352  & 0.1      \\ 
        $H3$ & 8.161  & 8.129  & 0.4      \\ 
        $H4$ & 9.971  & 9.875  & 1.0      \\ 
        $H5$ & 11.806 & 11.657 & 1.3      \\
        $H6$ & 13.605 & 13.421 & 1.7      \\
\end{tabular}
\label{table:radial}
\end{table}

After establishing the accuracy of the long-term
evolution of the GRHydro equations, we have examined the
eigenfrequencies of the radial pulsations of spherical
stars in {\it coupled} hydrodynamical and spacetime
evolutions. A Fourier transform of the evolution of the
radial velocity (for the same equilibrium model as the
one discussed before) is shown in
Fig.~\ref{fig:rho_fftN1.0.eps}.  Again in this case, we
have been able to identify several frequency peaks in the
Fourier spectrum with the normal mode frequencies
obtained with linear perturbation
techniques~\cite{Kokkotas01p}. A detailed comparison of
these frequencies is shown in
Table~\ref{table:radial_full}. The agreement is again
excellent. Note the rather large differences between the
frequencies shown in Tables~\ref{table:radial} and
\ref{table:radial_full}. The Cowling approximation is
rather inaccurate for the lowest radial
mode-frequencies~\cite{Yoshida01}, but is increasingly
more accurate for nonradial pulsations or for higher
frequencies~\cite{Yoshida01}.

\begin{table}[htb]
\caption{ Comparison of small-amplitude radial pulsation
frequencies obtained with the present 3D nonlinear
evolution code with linear perturbation mode frequencies,
in fully {\it coupled} evolutions. The equilibrium model
is a nonrotating $N=1$ relativistic polytrope with
$M/R=0.15$.}
\medskip
\begin{tabular}{cccc}
Mode & Present 3D code & Perturbation code & Relative
     Difference \\ 
        & (kHz) & (kHz) & (\%)           \\ 
\hline 
        $F$  & 1.450 & 1.442 & 0.6        \\ 
        $H1$ & 3.958 & 3.955 & 0.0        \\
        $H2$ & 5.935 & 5.916 & 0.3        \\ 
        $H3$ & 7.812 & 7.776 & 0.4        \\
\end{tabular}
\label{table:radial_full}
\end{table}

        All of the results discussed so far refer to
simulations involving stable relativistic
configurations. In the following section we consider
numerical evolutions of relativistic stars which are
initially in an unstable equilibrium.
 
\subsection{Migration of unstable configurations to the stable branch}
\label{migration}

\begin{figure}[htb]
\centerline{\psfig{file=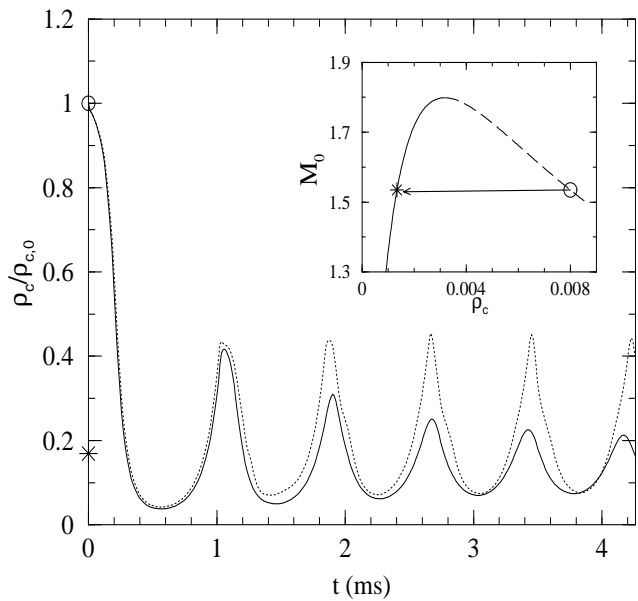,height=3.2in,width=3.3in}}
\caption{Evolution of the (normalized) central rest-mass
density $\rho_c$ during the migration of an unstable
relativistic star to a stable model with the same
rest-mass. When an adiabatic EOS is used (dotted line)
the difference in gravitational binding energy between
the unstable and stable models is periodically converted
in bulk kinetic energy through highly nonlinear, nearly
constant amplitude pulsations. In contrast, when an ideal
fluid EOS is used (solid line), the gravitational binding
energy is gradually converted into internal energy via
shock heating. As a result, the oscillations are damped
and the heated stable equilibrium model approaches a
central density slightly smaller than the rest-mass
density of a zero temperature star of the same rest-mass
(indicated by an asterisk on the left vertical axis).}
\label{fig:rho_migration.eps}
\end{figure}

	The numerical evolution of a nonrotating,
relativistic star in an equilibrium unstable to the
fundamental radial mode of pulsation is mainly determined
by the numerical truncation errors that cause it to
evolve away from its initial configuration. Depending on
the type of perturbation, the star can either collapse to
a black hole or expand and migrate to the stable branch
of the sequence of equilibrium models, reaching a new,
stable equilibrium configuration with approximately the
same rest-mass of the perturbed star. We have therefore
constructed a model of a $N=1, K=100$ polytropic star
with rest-mass $M_0=1.535\;M_\odot$ ($M=1.447\;M_\odot$)
and a central rest-mass density $\rho_c=8.0\times
10^{-3}$, which is larger than the central rest-mass
density of the maximum-mass stable model. The star is
therefore initially in an unstable equilibrium (see the
inset of Fig.~\ref{fig:rho_migration.eps}) and under the
perturbation introduced by the truncation error, it
expands, evolving rapidly to smaller central rest-mass
densities, until it reaches the stable branch of
equilibrium configurations. An analogous behavior has
been observed in numerical simulations of relativistic
boson stars~\cite{Seidel90b} (see also
Ref.~\cite{Hawley00} for recent numerical simulations of
expanding unstable boson stars).

	In a realistic astrophysical scenario, a stable
neutron star can accrete matter e.g. from a companion
star in a binary system or from infalling matter after
its formation in a supernova core-collapse. The star
would then secularly move towards larger central
densities along the stable branch of equilibrium
configurations, exceed the maximum-mass limit and
collapse to a black hole. No secular mechanism could
evolve the star to the unstable branch. In this respect,
the migration mechanism discussed here cannot occur in
practice. Nevertheless, it provides a consistent solution
of the initial value problem and represents an important
test of the accuracy of the code in a highly dynamical
and non-adiabatic evolution.  We use such an initial data
set to study large amplitude oscillations of relativistic
stars, which cannot be treated accurately by linear
perturbation theory.  Large amplitude oscillations about
a configuration on the stable branch could occur after a
supernova core-collapse~\cite{Dimmelmeier01} or after an
accretion-induced collapse of a white dwarf.  While the
actual set of quasi-normal modes excited will depend on
the excitation process, the ability to simulate large
amplitude oscillations is important.

\begin{figure}[htb]
\centerline{\psfig{file=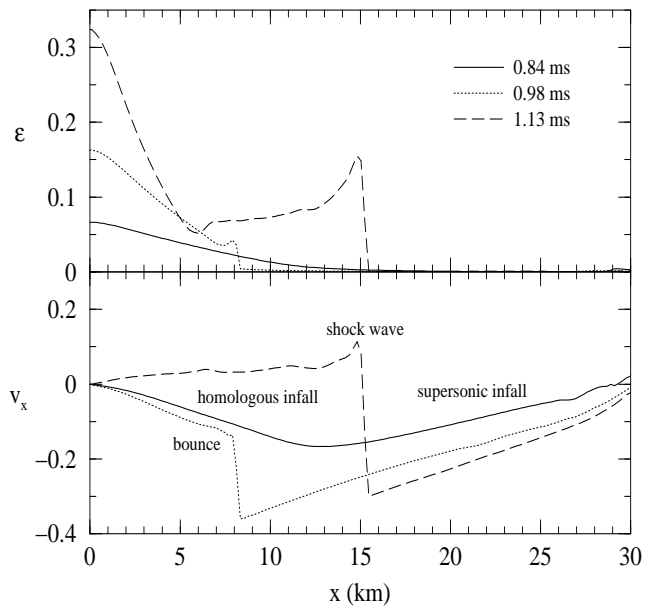,height=3.2in,width=3.3in}}
\caption{Shock formation in the outer core mantle, during the
first bounce at equilibrium densities of an unstable star, evolved
with an ideal fluid EOS. The top and bottom panels show
the internal energy $\epsilon$ and radial velocity $v_x$, respectively,
at three different times: the homologous infall phase, the inner
core bounce and the outwards shock propagation. The oscillations
of the inner core are damped by shock heating.}
\label{fig:shock_2in1}
\end{figure}

        Fig.~\ref{fig:rho_migration.eps} shows the
evolution of the central rest-mass density $\rho_c$
normalized to its initial value and up to a final time of
4.26 ms. On a very short dynamical timescale of 0.5 ms
the star has expanded and has its central density dropped
to about 3 \% of its initial central rest-mass
density. Note that this is less than the central
rest-mass density, $\rho_c=1.35\times 10^{-3}$, of the
stable model of same rest-mass, which is indicated with
an asterisk on the vertical axis of
Fig.~\ref{fig:rho_migration.eps}. During the rapid
decrease of the central rest-mass density, the star
acquires a large radial momentum. The star then enters a
phase of large amplitude radial oscillations about the
stable equilibrium model with the same rest-mass. Because
the unstable and stable models have rather different
degrees of compactness, the migration to the stable
branch will be accompanied by the release of a
significant amount of gravitational binding energy which
could either be converted to bulk kinetic energy or to
internal energy depending on the choice of EOS.

	In order to investigate both responses, we have
performed two different evolutions of the same initial
model. In the first case (the ``adiabatic EOS'' in
Fig.~\ref{fig:rho_migration.eps}), we have enforced the
adiabatic condition during the evolution, i.e. we have
assumed that the star remains at zero temperature
following an adiabatic EOS. As shown in
Fig.~\ref{fig:rho_migration.eps} with a dotted line, in
this case the star behaves like a compressed spring which
is allowed to expand, oscillating with a nearly constant
amplitude. This indicates that the star periodically
converts all of the excess gravitational binding energy
into the kinetic energy and vice versa. As the
oscillations are highly nonlinear, the restoring force is
weaker at higher densities than at lower densities and
the oscillations are therefore far from being sinusoidal.

	In the second case (the ``ideal fluid EOS'' in
Fig.~\ref{fig:rho_migration.eps}), we do not enforce the
abiabatic condition, but allow all of thermodynamic
variables to evolve in time. As a result, the
oscillations are gradually damped in time, while the star
oscillates around a central density close to that of a
stable star with the same rest-mass.

	The rapid decrease in the oscillation amplitude
is due to the dissipation of kinetic energy via shock
heating. At the end of the first expansion (i.e. at the
first minimum in Fig.~\ref{fig:rho_migration.eps}), the
star has expanded almost to the edge of the numerical
grid.  At this point, the outer parts of the initial star
have formed a low-density, outer-core mantle around the
high-density inner core and the star then starts to
contract. Fig.~\ref{fig:shock_2in1} shows with solid
lines the supersonic infall of the outer core mantle at
$t=0.84$ ms, while the inner core is contracting
homologously. After this ``point of last good homology",
the high-density inner core reaches its maximum infall
velocity and then starts slowing down. The infalling
low-density mantle forms a shock at the inner core/mantle
boundary (dotted lines at $t=0.98$ ms in
Fig.~\ref{fig:shock_2in1}). After the inner core bounces,
it expands and pressure waves at the inner core-mantle
boundary feed the shock wave with kinetic energy (dashed
lines at $t=1.13$ ms in Fig.~\ref{fig:shock_2in1}). In
this way, the shock wave is dissipating the initial
binding energy of the star so that the amplitude of the
central density oscillations decreases with time. The
above process is very similar to the core bounce in
neutron star formation (see, for instance, the
description in \cite{Moenchmeyer91}), except for the fact
that here the outer mantle is created during the first
rapid expansion from material of the initial unstable
star.

	As a result of the damping of the radial
oscillations, the star settles down, on a secular
timescale, to a stable equilibrium configuration with
central density somewhat smaller than the central density
of a stable star with same rest-mass as the initial
unstable star. This is because part of the matter of the
initial star forms a heated mantle around the inner core.

	The evolution shown in
Fig.~\ref{fig:rho_migration.eps} was obtained using a
resolution of $96^3$ gridpoints. Since the initial
unstable configuration is much more compact than the
final configuration, the boundaries of the computational
grid were placed at about $4.5\;r_s$. As a result, the
grid resolution of the initial configuration is rather
low, causing an additional, non-negligible deviation of
the average central rest-mass density of the pulsating
star away from the expected central rest-mass density of
the zero-temperature star of the same rest-mass.

	The evolution of the highly nonlinear and
nonadiabatic pulsations of a star when it settles down on
the stable branch, underlines the importance of evolving
all of the thermodynamic variables (including the
specific internal energy) and the importance of using
HRSC methods in order to resolve the formation and
evolution of shocks correctly. These capabilities of the
numerical code will be important in the correct
simulation of general relativistic astrophysical events
such as the merging of a neutron star binary system or
the formation of a neutron star in an accretion-induced
collapse of a white dwarf.

\subsection{Gravitational collapse of unstable configurations}
\label{collapse}

       As mentioned in the previous section, the
numerical scheme used in the hydrodynamical evolution is
such that it causes a nonrotating relativistic star in an
unstable equilibrium to expand and migrate to the
configuration of same rest-mass located on the stable
branch of equilibrium configurations. In order to study
the gravitational collapse to a black hole of an unstable
model we need to add to the initial model a
small radial perturbation in the rest-mass density
distribution. A very small perturbation of the order of
$\sim 1\%$ is sufficient and its radial dependence can be
simply given by $\cos(\pi r/2r_s)$, where $r$ is
coordinate distance from the center and $r_s$ its value
at the surface of the star. The addition of this small
perturbation dominates over the truncation error and
causes the star to collapse to a black hole. Note that
after the perturbation is added to the initial
equilibrium configuration, the constraint equations are
solved to provide initial data which is a solution to
the field equations~\cite{York79}.

\begin{figure}[htb]
\centerline{\psfig{file=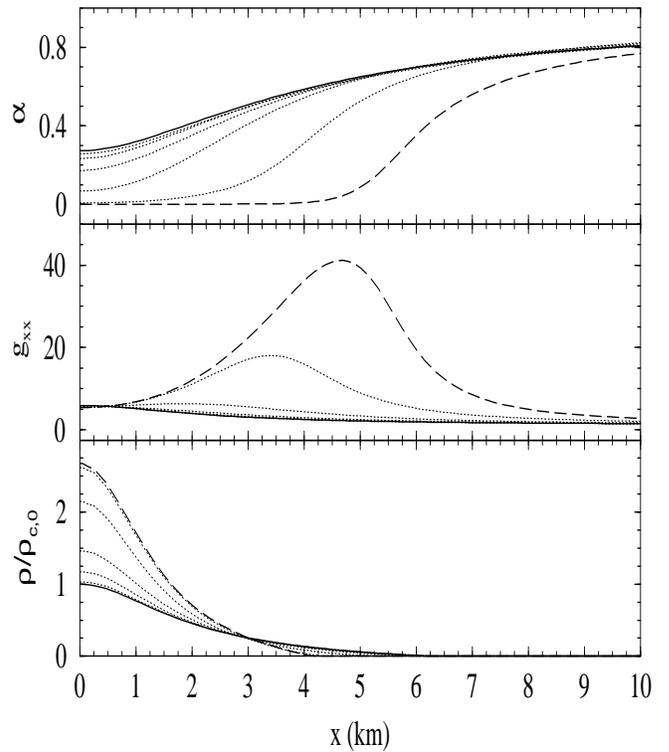,height=4.0in,width=3.4in}}
\caption{Profiles along the $x$-axis of representative
metric and fluid quantities during the gravitational
collapse to a black hole of an unstable $N=1$,
$\rho_c=8.0 \times 10^{-3}$ relativistic polytrope
showing different snapshots of the time evolution. The
top, medium and bottom panels show the evolution of the
lapse function, of the $g_{xx}$ metric component, and of
normalized rest-mass density, respectively. The thick
solid and dashed lines indicate the initial and final
(\hbox{$t=0.29$ ms}) profiles.  Intermediate profiles,
indicated by thin dotted ashed lines, are shown every
0.049 ms. }
\label{fig:collapse.eps}
\end{figure}

        The (forced) collapse to a black hole of an
unstable spherical relativistic star is shown in
Fig.~\ref{fig:collapse.eps} for a simulation with $128^3$
gridpoints in octant symmetry, using Roe's solver and an
ideal fluid EOS. The figure shows the profiles along the
$x$-axis of the lapse function (top panel), of the
$g_{xx}$ metric component (middle panel) and of the
normalized rest-mass density (bottom panel).  Different
lines refer to different times of the evolution, with the
thick solid line in each panel indicating the initial
profile and with the thick dashed line corresponding to
the final timeslice at \hbox{$t=0.29$ ms}; intermediate
times (shown every 0.049 ms) are indicated with dotted
lines. The evolution of the lapse function shows the
characteristic ``collapse of the lapse'', a distinctive
feature of black hole formation. The evolution of the
$g_{xx}$ metric component and of the rest-mass density
also clearly exhibit features typical of black hole
formation, such as the large peak developing in $g_{xx}$,
or the continuous increase in the central rest-mass
density.
 
\begin{figure}[htb]
\centerline{\psfig{file=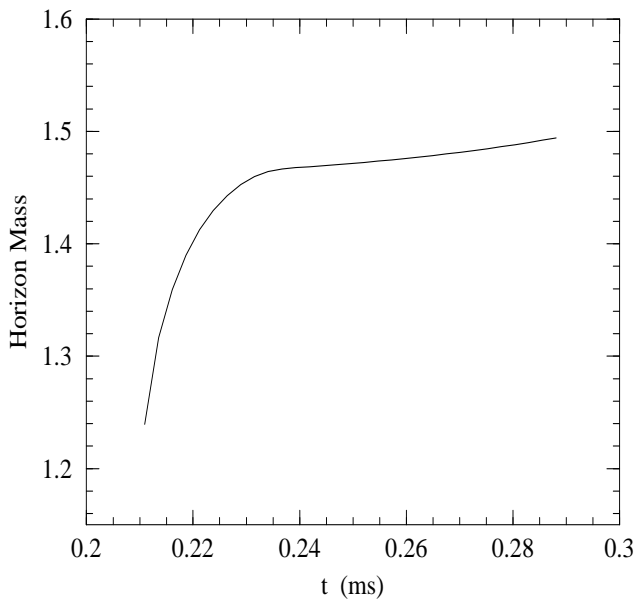,height=3.2in,width=3.3in}}
\caption{Horizon Mass as a function of time. A black hole
is formed at \hbox{$t=0.21$ ms} and the horizon mass then
starts to increase as a result of accretion.}
\label{fig:horizon_mass.eps}
\end{figure}

        While the collapse of the lapse is a good
indication of the formation of a black hole, the
formation of an apparent horizon (the outermost of the
trapped surfaces) in the course of the simulation is an
unambiguous signature of black hole formation.  An
apparent horizon finder based on the fast-flow
algorithm~\cite{Gundlach97a} was used to detect the
appearance of horizons, and to calculate the horizon
mass. This apparent horizon finder, and its validation,
is described in Ref.~\cite{Alcubierre98b}.

	Fig.~\ref{fig:horizon_mass.eps} shows the
evolution of the horizon mass as a function of
time. Initially there is no horizon. At a time $t=0.21$
ms a black hole forms and an apparent horizon appears. As
the remaining stellar material continues to accrete onto
the newly formed black hole its horizon mass increases,
finally levelling off, until about $t=0.27$ ms. The
subsequent growth of the horizon mass is the result of
the increasing error due to grid stretching - the radial
metric function develops a sharp peak which cannot be
resolved adequately.

\section{Rapidly Rotating Relativistic Stars}
\label{RRNSs}

\subsection{Stationary equilibrium models}

        The long-term evolution of rapidly rotating,
stable equilibrium relativistic stars represents a much
more demanding test for a numerical code. In this case,
in fact, the use of a non-zero shift vector is strictly
necessary and this, in turn, involves the testing of
parts of the code that are not involved in the evolution
of a non-rotating stellar model. The initial data used
here are numerical solutions describing general
relativistic stationary and axisymmetric equilibrium
models rotating uniformly with angular velocity
$\Omega$. The models are constructed with the {\tt rns}
code \cite{Stergioulas95,Nozawa98} (see
Ref.~\cite{Stergioulas98} for a recent review of rotating
stars in relativity) which provides the four metric
potentials $\nu$, $B$, $\mu$, and $\omega$ needed to
describe the spacetime with line element
\begin{eqnarray} 
ds^2 & =& -e^{2 \nu } dt^2 + B^2 e^{-2 \nu} r^2 \sin^2
        \theta ( d \phi - \omega dt )^2 \nonumber \\ & \
        & + e^{2 \mu }(dr^2+r^2 d \theta^2).
\end{eqnarray}
In the nonrotating limit, the above metric reduces to the
metric of a static, spherically symmetric spacetime in
isotropic coordinates.  A rotating model is uniquely
determined upon specification of the EOS and two
parameters, such as the central rest-mass density and the
ratio of the polar to the equatorial coordinate radii
(axes ratio).

\begin{figure}[htb]
\centerline{\psfig{file=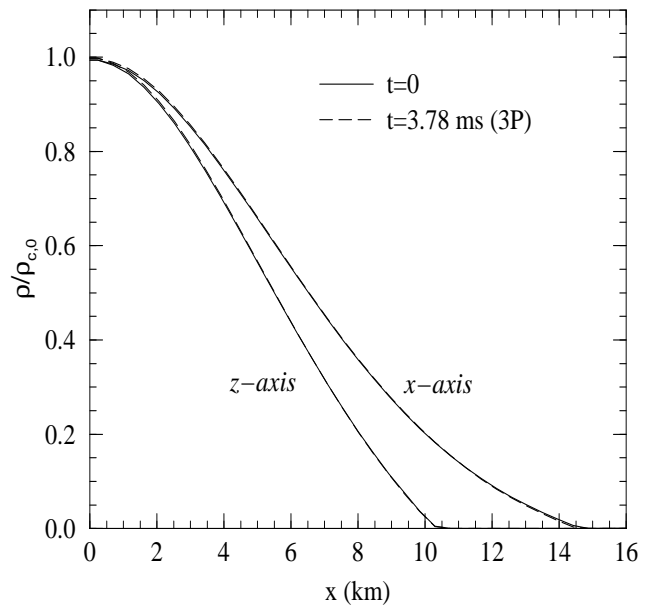,height=3.2in,width=3.3in}}
\caption{ Profiles of the (normalized) rest-mass density
along the $x$-axis and $z$-axis at two coordinate times,
$t=0$ (solid lines) and \hbox{$t=3.78$ ms} (dashed
lines), corresponding to three rotational periods
($P$). The star is a $N=1$, $\rho_c=1.28\times 10^{-3}$
polytrope rotating at $92\%$ of the mass-shedding
limit. The simulation has been performed only in the
volume above the $(x,y)$ plane which is covered with
$129\times 129\times 66$ gridpoints.}
\label{fig:rho_rot.eps}
\end{figure}

        Using the standard Jacobian transformations
between the spherical polar coordinates $(r,\theta,\phi)$
and the Cartesian coordinates $(x,y,z)$, the initial data
for a rotating star are transformed to Cartesian
coordinates. Convergence tests of the initial data on the
Cartesian grid at various resolutions, show that the
Hamiltonian and momentum constraints converge at
second-order everywhere except at the surface of the
star, where some high-frequency noise is present. This
noise is due to Gibbs phenomena at the surface of the
star, which are inherent to the method~\cite{Komatsu89}
used in the construction of the 2D initial data (see the
relevant discussion in Ref.~\cite{Nozawa98}).  To our
knowledge, all currently available methods for
constructing initial data describing rotating
relativistic stars suffer from some kind of Gibbs
phenomena at the surface of the star, with the only
exception being a recent multi-domain spectral method that
uses surface-adapted coordinates~\cite{Bonazzola98a}. The
high-frequency noise does not appear to affect the
long-term evolution of the initial data at the grid
resolutions employed in our simulations. The evolution is
carried out up to several rotational periods, using the
shift 3-vector obtained from the solution of the
stationary problem, which we do not evolve in time.

\begin{figure}[htb]
\centerline{\psfig{file=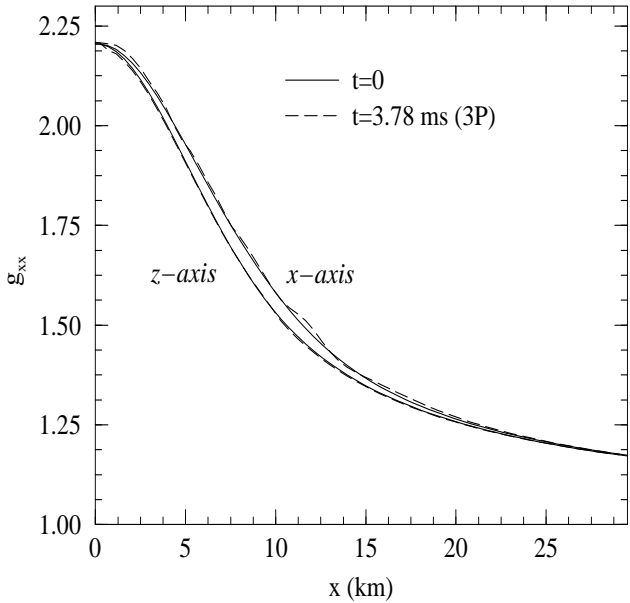,height=3.2in,width=3.3in}}
\caption{ Profile of the metric component $g_{xx}$ along
the $x$-axis and $z$-axis at two different coordinate times, for the
same evolution shown in Fig.~\ref{fig:rho_rot.eps}.}
\label{fig:gxx_rot.eps}
\end{figure}

        We have evolved models at various rotation rates
and for several polytropic EOS, all showing similar
long-term behaviour and convergence.  Hereafter we will
focus on a $N=1$ polytropic model, rotating at $92\%$ of
the allowed mass-shedding limit for a uniformly rotating
star with the same central rest-mass density. In
particular, we have chosen a stellar model with the same
central rest-mass density as the nonrotating model of
Section~\ref{sphstar} and which is significantly
flattened by the rapid rotation (the polar coordinate
radius is only 70 \% of the equatorial coordinate
radius).

        Similarly to what is observed in the numerical
evolution of nonrotating stars, the truncation errors
trigger, in a rapidly rotating star, oscillations that
are quasi-radial. As a result, the rotating star pulsates
mainly in its fundamental quasi-radial mode and, during
the long-term evolution, its central rest-mass density
drifts towards higher values. Also in this case, both the
amplitude of the pulsations and the central density
growth rate converge to zero at nearly second-order with
increasing grid resolution.

        Our simulations have been performed only in the
volume above the $(x,y)$ plane which is covered with
$129\times129\times 66$ gridpoints. At such grid
resolutions, we have been able to evolve a stationary
rapidly rotating relativistic star for three complete
rotational periods, before the numerical solution departs
noticeably from the initial configuration. Note that much
longer evolution times (more than an order of magnitude
longer and essentially limited by the time available) can
be achieved if the spacetime is held fixed and only the
hydrodynamical equations in a curved background are
evolved. This has been demonstrated recently in
Ref.~\cite{Stergioulas01}, with a code based on the one
used in the present paper and in which a third-order
Piecewise Parabolic Method (PPM)~\cite{Colella84} was
used for the hydrodynamical evolution and applied to the
study of nonlinear $r$-modes in rapidly rotating
relativistic stars and the occurrence of differential of
a kinematical differential
rotation~\cite{Rezzollaetal2000} (see
Ref.~\cite{Andersson01,Friedman01} for a recent review on
the $r$-mode instability). While our current second-order
TVD method with the MC limiter is not as accurate (for
the same grid resolution) as the third-order PPM method,
it has, nevertheless, a very good accuracy, significantly
better than that of the minmod limiter.

\begin{figure}[htb]
\centerline{\psfig{file=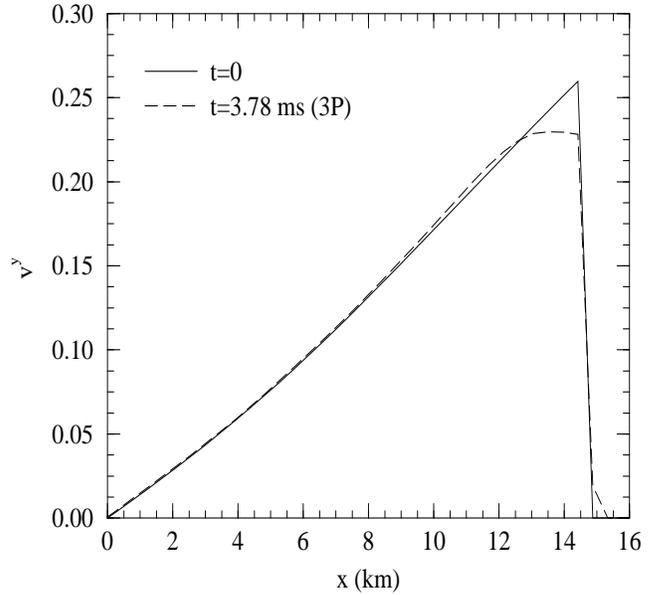,height=3.2in,width=3.3in}}
\caption{ The velocity component $v^y$ along the $x$-axis
at two different coordinate times, for the same
evolution as in Fig.~\ref{fig:rho_rot.eps}.}
\label{fig:velx_yl_rot.eps}
\end{figure}

Results of our simulations of rapidly-rotating stars are
plotted in
Figs.~\ref{fig:rho_rot.eps}-\ref{fig:velx_yl_rot.eps}.
In particular, Fig.~\ref{fig:rho_rot.eps} shows the
(normalized) rest-mass density along the $x$ and $z$ axes
at two coordinate times, $t=0$ (solid lines) and
\hbox{$t=3.78$ ms} (dashed lines), with the latter
corresponding to three rotational periods. The outer
boundary of the grid is placed at about twice the
equatorial radius. After three rotational periods, the
rest-mass density profile is still very close to the
initial one. Similarly, Fig.~\ref{fig:gxx_rot.eps} shows
the metric component $g_{xx}$ along the $x$ and $z$ axes
at the same coordinate times of
Fig.~\ref{fig:rho_rot.eps}. Again, the change in $g_{xx}$
is minimal and only near the stellar surface can one
observe a noticeable difference (the error there grows
faster, due to the Gibbs phenomenon in the initial data).

	Besides triggering the appearance of quasi-radial
pulsations and the secular increase in the central
rest-mass density, the truncation errors also induce the
formation of a local maximum at the stellar surface for
the evolved ``momentum'' variable $\tilde{S_j}$ [cf.
Eq.~(\ref{eq:evolvedvar})]. The existence of this local
extremum reduces, at the surface of the rotating star,
the order of our TVD schemes to first-order only. As a
result, the angular momentum profile at the surface
gradually drifts away from the initial uniformly rotating
one, with the rate of convergence of this drift being
only first-order with increasing grid resolution. We
emphasize, however, that this is only a local effect:
everywhere else inside the star, the angular momentum
evolution is second-order accurate.
Fig.~\ref{fig:velx_yl_rot.eps} shows the velocity
component $v^y$ along the $x$-axis at the same coordinate
times of Fig.~\ref{fig:rho_rot.eps} and
\ref{fig:gxx_rot.eps}.  Alternative evolution schemes
based on third-order methods have been shown to have a
smaller truncation error at the surface of the star, both
for 2D and 3D evolutions of the same initial
data~\cite{Font99,Stergioulas01}, at least in the Cowling
approximation.

	Note that plotting the velocity profile as in
Fig.~\ref{fig:velx_yl_rot.eps} allows one to ascertain the
accuracy in the preservation of the velocity
field. Isocontours or vector plots of the velocity field
can, in fact, easily mask the secular evolution shown in
Fig.~\ref{fig:velx_yl_rot.eps}. We also note that the
variable evolved in the code is not the rotational
velocity, but a corresponding momentum component which
depends on the local rest-mass
[cf. Eq.(~\ref{eq:evolvedvar})]. The error in the
rotational velocity near the surface is therefore also
influenced by the small value of the rest-mass density in
that region.

\subsection{Quasi-radial modes of rapidly rotating
relativistic stars}

	The quasi-radial pulsations of rotating neutron
stars are a potential source of detectable gravitational
waves and could be excited in various astrophysical
scenarios, such as a rotating core-collapse, a core-quake
in a rotating neutron star (due to a large
phase-transition in the equation of state) or the
formation of a high-mass neutron star in a binary neutron
star merger. An observational detection of such
pulsations would yield valuable information about the
equation of state of relativistic
stars~\cite{Kokkotas01}. So far, however, the
quasi-radial modes of rotating relativistic stars have
been studied only under simplifying assumptions such as
in the slow-rotation
approximation~\cite{Hartle75,Datta98} or in the
relativistic Cowling
approximation~\cite{Yoshida01,Font01}. The spectrum of
quasi-radial pulsations in full General Relativity has
not been solved to date with perturbation techniques (see
Ref.~\cite{Stergioulas98} for a recent review of the
subject).

        In this section we take a step forward in the
solution of this long standing problem in the physics of
relativistics stars and obtain the first mode-frequencies
of rotating stars in full General Relativity and rapid
rotation. As done in Section~\ref{rp} for the radial
pulsation of nonrotating stars, we take advantage of the
very small numerical viscosity of our code to extract
physically relevant information from the quasi-radial
perturbations induced by truncation errors. The ability
to do so demonstrates that our current numerical methods
are mature enough to obtain answers to new problems in
the physics of relativistics stars.

\begin{table}[htb]
\caption{Comparison of small-amplitude quasi-radial
pulsation frequencies obtained with the present 3D code
in {\it fixed spacetime}, with frequencies obtained with
an independent 2D code. The equilibrium model is a $N=1$
relativistic polytrope rotating at 92\% of the
mass-shedding limit.}
\medskip
\begin{tabular}{cccc}
Mode & Present 3D code & 2D code & Relative Difference  \\
     &  (kHz)          &  (kHz)            &  (\%) \\
\hline
        $F$  & 2.468 & 2.456 & 0.5        \\
        $H1$ & 4.344 & 4.357 & 0.3        \\
        $H2$ & 6.250 & 6.270 & 0.3        \\
\end{tabular}
\label{table:quasi-radial_Cowling}
\end{table}

\begin{table}[htb]
\caption{Quasi-radial pulsation frequencies for a
sequence of rotating $N=1$ polytropes with rotation rates
up to 97\% of the mass-shedding limit. The frequencies of
the fundamental mode $F$ and of the first overtone $H1$
are computed from {\it coupled} hydrodynamical and
spacetime evolutions. The ratio of polar $r_p$ to
equatorial $r_e$ coordinate radii of the rotating models
is also shown.}
\medskip
\begin{tabular}{cccc}
$r_p/r_e$ & $\Omega/\Omega_K$ & $F$ (kHz)& $H1$ (kHz)	\\
\hline  
1.000 & 0.000 & 1.450 & 3.958 \\
0.950 & 0.407 & 1.411 & 3.852 \\
0.850 & 0.692 & 1.350 & 3.867 \\
0.825 & 0.789 & 1.329 & 3.894 \\
0.775 & 0.830 & 1.287 & 3.953 \\
0.750 & 0.867 & 1.265 & 4.031 \\
0.725 & 0.899 & 1.245 & 3.974 \\
0.700 & 0.929 & 1.247 & 3.887 \\
0.675 & 0.953 & 1.209 & 3.874 \\
0.650 & 0.974 & 1.195 & 3.717 
\end{tabular}
\label{table:quasi-radial_full}
\end{table}

	Following the approach outlined in
Section~\ref{rp}, we have first computed the quasi-radial
mode frequencies from numerical evolutions of the GRHydro
equations in a {\it fixed spacetime} evolution in order
to compare with recent results coming from an independent
2D nonlinear evolution code~\cite{Font01}. Table
\ref{table:quasi-radial_Cowling} shows the comparison of
between the eigenfrequencies computed in the Cowling
approximation with the 2D code for the equilibrium model
of the previous Section. Note that the newly obtained
frequencies differ by less than 0.5\%, verifying that our
code can accurately reproduce them.

\begin{figure}[htb]
\centerline{\psfig{file=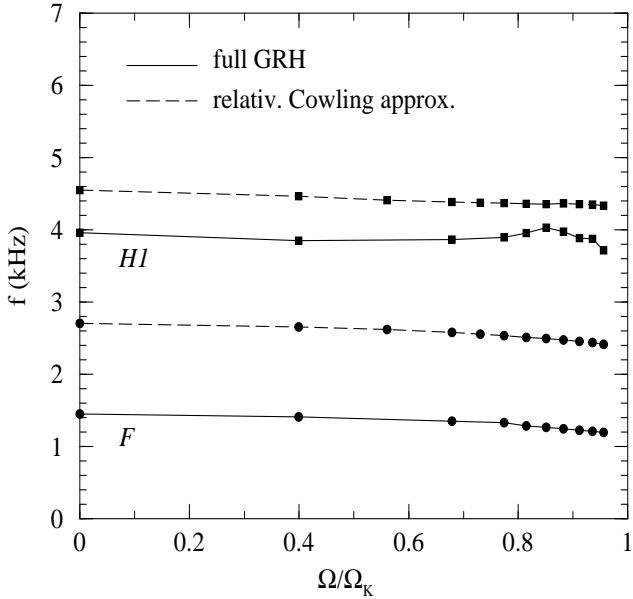,height=3.2in,width=3.3in}}
\caption{Quasi-radial pulsation frequencies for a
sequence of rotating $N=1$ polytropes and a number of
different rotation rates. The frequencies of the
fundamental mode $F$ (filled squares) and of the first
overtone $H1$ (filled circles) are computed from {\it
coupled} hydrodynamical and spacetime evolutions (solid
lines). The sequences are also compared with the
corresponding results obtained from computations in the
relativistic Cowling approximation.}
\label{fig:freqs.eps}
\end{figure}

Next, we have computed the quasi-radial frequencies in
{\it coupled} hydrodynamical and spacetime evolutions for
rapidly rotating stars. As mentioned before, this is a
novel study and the results obtained cannot be 
compared with corresponding results in the literature.
To study this, we have carried out two types of
analysis. Firstly, we have followed the same procedure
used in the case of a nonrotating star case and obtained
the normalized frequency spectrum of oscillations induced
by the truncation errors. Secondly, we have computed the
frequency spectrum of oscillations triggered by a small
but specified perturbation. More precisely, we have
introduced the same radial perturbation in the rest-mass
density used in Sect.~\ref{collapse} to induce collapse:
i.e. $A\cos(\pi r/2r_p)$, where $A=0.02$, $r$ is
coordinate distance from the center, and $r_p$ is the
radial coordinate position of the poles. When compared,
the results of the two treatments indicate that the
fundamental mode frequency agrees to within 2\%, while the
H1 mode near the mass-shedding limit is probably accurate
to several percent only (at this resolution).

	To study quasi-radial modes of rapidly rotating
relativistic stars we have built a sequence of models
having the same grid resolution, the same equation of
state and central rest-mass density used in the previous
section, varying only the rotation rate $\Omega$. The
sequence starts with a nonrotating star and terminates
with a star at 97\% of the maximum allowed rotational
frequency $\Omega_K = 0.5363 \times 10^4$ s$^{-1}$ for
uniformly rotating stars (mass-shedding limit). The
results of these simulations are reported in
Table~\ref{table:quasi-radial_full} and shown in
Fig.~\ref{fig:freqs.eps}, where the frequencies of the
lowest two quasi-radial modes are shown.  Interestingly,
the fundamental mode-frequencies (solid lines) and their
first overtones (dashed lines) show a dependence on the
increased rotation which is similar to the one observed
for the corresponding frequencies in the Cowling
approximation~\cite{Font01}.

	In particular, the $F$-mode frequency decreases
monotonically as the maximum rotation rate is
approached. Near the mass-shedding limit, the frequency
is 18\% smaller than the frequency of the nonrotating
star. The difference between the $F$-mode frequency
computed here and the corresponding result in the Cowling
approximation is nearly constant. Thus, one can construct
an approximate empirical relation for the fundamental
quasi-radial frequency of rapidly rotating stars, using
only the corresponding frequency in the Cowling
approximation, $F_{\rm Cowling}$ and the frequency of the
fundamental radial mode in the nonrotating limit,
$F_{\Omega=0}$. For the particular sequence shown above,
the empirical relation reads
\begin{equation}
F=\left(F_{\rm Cowling}-1.246\right)\ {\rm kHz}\ ,
\end{equation}
and yields the correct frequencies with an accuracy of
better than 2\% for the most rapidly rotating model. More
generally, if $F_{\rm Cowling, \Omega=0}$ is the
frequency of the fundamental radial mode in the Cowling
approximation, then the empirical relation can be written as
\begin{equation}
F= F_{\Omega=0} + F_{\rm Cowling}- 
	F_{\rm Cowling,\Omega=0} \ . 
\end{equation}
Such an empirical relation is very useful, as it allows
one to obtain a good estimate of the fundamental
quasi-radial mode frequency of rapidly rotating stars by
solving the hydrodynamical problem in a fixed spacetime,
rather than solving the much more expensive evolution
problem in which the spacetime and the hydrodynamics are
coupled.

	The frequency of the $H1$ mode shows a
non-monotonic decrease as the mass-shedding limit is
approached, departing from the behavior obtained in the
Cowling approximation. The oscillations in the frequency
at larger rotation rates could be due to ``avoided
crossings'' with frequencies of other modes of
oscillation (We recall that is referred to as ``avoided
crossing'' the typical behaviour shown by two
eigenfrequency curves which approach smoothly but then
depart from each other without crossing. At the point of
closest approach, the properties of the modes on each
sequence are exchanged~\cite{unnoetal89}.). Similar
avoided crossings have been observed also in the Cowling
approximation for higher overtones and near the
mass-shedding limit (see
Refs.~\cite{Yoshida01,Font01}). Our results indicate
therefore that the avoided crossings in a sequence of
relativistic rotating stars occur for smaller rotation
rates than predicted by the Cowling approximation. This
increases the importance of avoided crossings and makes
the frequency spectrum in rapidly rotating stars more
complex than previously thought.

\section{Gravitational Waves from a Pulsating Star}
\label{gws}

        The ability to extract gravitational wave
information from simulations of relativistic compact
objects is an important feature of any 3D General
Relativistic hydrodynamics code. To assess the ability of
our code to extract self-consistent and accurate
gravitational waveforms we have excited simple
quadrupolar perturbations in our standard spherical $N=1$
polytrope. In particular, on the basis of the angular
behavior of the $\ell=2$, $f$-mode in linear perturbation
theory, we have introduced in the initial model a
perturbation in the velocity of the form
\begin{equation}
u_\theta (t=0) = A \sin{(\pi r/r_s)} \sin \theta \cos \theta.  
\end{equation}
where $A=0.02$ is the amplitude of the perturbation and
$r_s$ is the coordinate radius of the star.

       Following York~\cite{York79}, we have then
constructed the initial data for the perturbed model by
solving the constraint equations for the unperturbed
model with added perturbations and then proceeded 
to evolve this solution in time.
As a response to the initial
perturbations, the star has started a series of periodic
oscillations, mainly in the fundamental quadrupolar mode
of oscillation. Other, higher-order modes are also
excited (and observed) but these are several orders of
magnitude smaller and play no dynamical role.

\begin{figure}[htb]
\centerline{\psfig{file=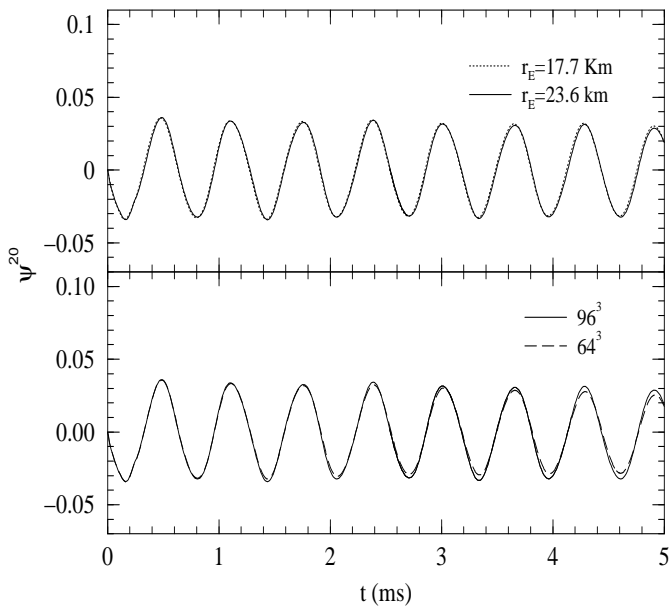,height=3.2in,width=3.3in}}
\caption{Gravitation-wave extraction ($\psi^{(20)}$) of a
perturbed nonrotating relativistic star, pulsating mainly
in the fundamental quadrupolar mode. The top panel shows
the rescaled waverforms as extracted at $r_{_{\rm
E}}=17.7$ km (dotted line) and at $r_{_{\rm E}}=23.6$ km
(solid line).  The lower panel, on the other hand, shows
the amplitude of $\psi^{20}$ extracted at $r_{_{\rm
E}}=23.6$ km for two different resolutions of $64^3$
(dashed line) and $96^3$ gridpoints (solid line),
respectively.}
\label{fig:psi2.eps}
\end{figure}

        As a consequence of the time-varying mass
quadrupolar triggered by the oscillations, the perturbed
star emits gravitational waves, which are extracted
through a perturbative technique discussed in detail in
Refs.~\cite{Camarda97c,Rupright98,Rezzolla99a}, and in
which the Zerilli function is expanded in terms of
spherical harmonics with each component being the
solution of an ordinary differential equation.

        We plot, in Fig.~\ref{fig:psi2.eps}, the
$\ell=2,m=0$ component of the Zerilli function
$\psi^{20}$. The upper panel, in particular, shows the
waverforms as extracted at $r_{_{\rm E}}=17.7$ km (dotted
line) and at $r_{_{\rm E}}=23.6$ km (solid line)
respectively, with the first having been rescaled as
$r^{-3/2}$ to allow a comparison. The very good agreement
between the two waveforms is an indication that the
gravitational waves have reached their asymptotic
waveform. The lower panel, on the other hand, shows the
amplitude of $\psi^{20}$ extracted at $r_{_{\rm E}}=23.6$
km for two different resolutions of $64^3$ (dashed line)
and $96^3$ gridpoints (solid line), respectively. Note
that the gravitational wave signal converges to a
constant amplitude, as the true gravitational-wave
damping timescale for this mode is several orders of
magnitude larger than the total evolution time shown. The
small decrease in the amplitude observed in our numerical
evolution is thus entirely due to the effective numerical
viscosity of our scheme. At a resolution of $96^3$
gridpoints, the effective numerical viscosity is
sufficiently low to allow for a quantitative study of
gravitational waves from pulsating stars over a timescale
of many dynamical times (the largest relative numerical
error estimated on the basis of the simulations presented
in Fig.~\ref{fig:psi2.eps} is 3.3\%).

        As done in the previous sections, we have
compared the frequencies derived from our numerical
simulations with those obtained from perturbative
techniques for $f$-mode oscillations of $N=1$
polytropes. Again in this case the
comparison has revealed a very good agreement between the
two approaches. As one would expect, the dominant
frequency of the gravitational waves we extract (1.56 kHz) agrees
with the fundamental quadrupole $f$-mode frequency of the star
(1.58 kHz) ~\cite{Kokkotas01p} to within 1.3 \% (at a 
resolution of $96^3$ grid-points).

\section{Conclusions}                                                   %
\label{sec:conclusion}                                                  %

        We have presented results obtained with a 3D
general relativistic code GR\_Astro in a
comprehensive study of the long-term dynamics of
relativistic stars. The code has been built by the
Washington University/Albert Einstein Institute
collaboration for the NASA Neutron Star Grand Challenge
Project~\cite{GR3D} and is based on the Cactus
Computational Toolkit~\cite{Cactusweb}. The simulations
reported here have benefited from several new numerical strategies
that have been implemented in the code and that concern
both the evolution of the field equations and the
solution of the hydrodynamical equations. In addition to
the features of the code discussed in paper I, the
present version of the code can construct various type of
initial data representing spherical and rapidly rotating
relativistic stars, extract gravitational waves produced
during the simulations and track the presence of an
apparent horizon when formed.
	
	All of these improvements have allowed tests and
performances well superior to those reported in the
companion paper I. With this improved setup we have shown
that our code is able to succesfully pass stringent
long-term evolution tests, such as the evolution of both,
static and rapidly rotating, stationary configurations.
We have also considered the evolution of relativistic
stars unstable to either gravitational collapse or
expansion. In particular, we have shown that unstable
relativistic stars can, in the course of a numerical
evolution, expand and migrate to the stable branch of
equilibrium configurations. As an application of this
property, we have studied the large-amplitude nonlinear
pulsations produced by the migration. Nonlinear
oscillations are expected to accompany the formation of a
proto-neutron star after a supernova core-collapse or
after an accretion-induced collapse of a white dwarf.

	Particularly significant for their astrophysical
application, we have investigated the pulsations of both
rapidly rotating and nonrotating relativistic stars and
compared the computed frequencies of radial, quasi-radial
and quadrupolar oscillations with the frequencies
obtained from perturbative methods or from axisymmetric
nonlinear evolutions. We have shown that our code
reproduces these results with excellent accuracy. As a
particularly relevant result, we have obtained the first
mode-frequencies of rotating stars in full general
relativity and rapid rotation. A long standing problem,
such frequencies had not been obtained so far by
other methods.

        In our view the results discussed in this paper
have a double significance. Firstly, they establish the
accuracy and reliability of the numerical techniques
employed in our code and which, to the best of our
knowledge, represent the most accurate long-term 3D
evolutions of relativistic stars available to
date. Secondly, they show that our current numerical
methods are mature enough for obtaining answers to new
problems in the physics of relativistic stars.

\acknowledgements                                                       %

It is a pleasure to thank K.D. Kokkotas for many
discussions and for providing us with the linear
perturbation frequencies. We have also benefited from
many discussions with M. Alcubierre, S. Bonazzola,
B. Br{\"u}gman, J.M. Ib{\'a}nez, J. Miller, M. Shibata,
K. Uryu, and Shin Yoshida.

\noindent The simulations in this paper have made use of
code components developed by several authors. In what
follows we report the names of the different components,
their use and their main author. BAM (elliptic equation
solver) B. Br{\" u}gman; AH-FINDER (apparent horizon
finder) M. Alcubierre; CONF-ADM (evolution scheme for the
field equations) and MAHC (evolution scheme for the GRHydro
equations) M. Miller; PRIM-SOL (solver for the
hydrodynamical primitive variables) P. Gressman; RNS-ID
(initial data solver for rotating and perturbed relativistic stars)
N. Stergioulas; EXTRACT (gravitational wave analysis)
G. Allen. The application code is built on the CACTUS
Computational Toolkit written by P. Walker {\it et al}
(version 3) and T. Goodale {\it et al} (version 4).

\noindent Financial support for this research has been
provided by the NSF KDI Astrophysics Simulation
Collaboratory (ASC) project (Phy 99-79985), NASA Earth
and Space Science Neutron Star Grand Challenge Project
(NCCS-153), NSF NRAC Project Computational General Relativistic
Astrophysics (93S025), and the EU Programme
"Improving the Human Research Potential and the
Socio-Economic Knowledge Base" (Research Training Network
Contract HPRN-CT-2000-00137).



\begin{thebibliography}{100}
\bibitem{GR3D} The code and its documentation can be
found at {\footnotesize wugrav.wustl.edu/Codes/GR3D}
\end{thebibliography}


\end{document}